\documentclass[pra,twocolumn,nofootinbib]{revtex4-1}

\usepackage{graphicx}
\usepackage{bm, mathbbol}
\usepackage{hyperref}
\usepackage{amsmath,amssymb,amsfonts}
\usepackage{braket}

\begin{document}

\title{Macroscopic boundary effects in the one-dimensional extended Bose-Hubbard model}

\author{Sebastian Stumper}
\affiliation{Physikalisches Institut, Albert-Ludwigs-Universit\"at Freiburg, Hermann-Herder-Stra\ss e 3, 79104 Freiburg, Germany}
\email{sebastian.stumper@physik.uni-freiburg.de}

\author{Junichi Okamoto}%
\affiliation{Physikalisches Institut, Albert-Ludwigs-Universit\"at Freiburg, Hermann-Herder-Stra\ss e 3, 79104 Freiburg, Germany}
\date{\today}

\begin{abstract}
We study the effect of different open boundary conditions on the insulating ground states of the one-dimensional extended Bose-Hubbard model  at and near unit filling.  To this end, we employ the density matrix renormalization group method with system sizes up to 250 sites. To characterize the system, various order parameters and entanglement entropies are calculated. When opposite edge potentials are added to the two ends of the chain, the inversion symmetry is explicitly broken, and the regular bulk phases appear. On the other hand, simple open boundary conditions often exhibit non-degenerate ground states with a domain wall in the middle of the chain, which induces a sign-flip of an order parameter. Such a domain wall can lead to an algebraic behavior of the off-diagonals of the single particle density matrix. We show that this algebraic behavior adds only a finite contribution to the entanglement entropy, which does not diverge as the system size increases. Therefore, it is not an indication of a superfluid phase. We confirm this picture by analytical calculations based on an effective Hamiltonian for a domain wall.
\end{abstract}

\maketitle

\section{\label{sec:intro}Introduction}
Cold atoms in optical lattices provide flexible experimental settings for implementing a large variety of many-body Hamiltonians. Due to the controllability of model parameters, a wealth of quantum phases has been created such as superfluids, Mott insulators, or Luttinger liquids \cite{Bloch2008}. The cold atom setups can also serve as quantum simulators for solid-state systems~\cite{Lewenstein2007,Jaksch2005}. Although the on-site interaction in an optical lattice is usually dominant due to short-ranged $s$-wave interaction, it is possible to introduce long-ranged interactions, for instance by using dipolar magnetic atoms or Rydberg atoms~\cite{Lahaye2009,Baier2016}.

One of the most studied models for long-ranged interactions includes an on-site and a nearest-neighbor interaction term. For bosonic particles it is called an extended Bose-Hubbard model (EBHM). In the strongly interacting regime, the model hosts various insulating phases such as a Mott insulator (MI) or a density wave (DW) phase~\cite{Kuehner2000}. In one-dimension (1D), the Haldane insulator (HI), an analog of the Haldane phase in antiferromagnetic spin-$1$ chains, appears~\cite{DallaTorre2006,Berg2008, Ejima2014PRL}. This is a symmetry-protected topological phase~\cite{Pollmann2010}, which cannot be distinguished from the MI based on any local order parameter. Rather the number fluctuations in the ground states of MI and HI display different patterns, which are detected by non-local correlation functions. The HI is expected to exhibit fractionally charged edge states, which makes it a particularly appealing object to study due to possible applications in quantum computing. Understanding and controlling the edge states is thus an important problem.

In this work, we propose to use boundary conditions as another means of controlling insulating ground states of the 1D EBHM at or near unit filling. We consider various open boundary conditions, which we expect to be experimentally realizable. For example, local edge potentials can be accomplished by using optical tweezers~\cite{Muldoon2012}. The dependencies on the boundary conditions of order parameters and of the entanglement entropies are numerically studied by a density-matrix renormalization group (DMRG) method based on matrix product states (MPS)~\cite{openMPS}. When opposite edge potentials are applied, the inversion symmetry of the model is explicitly broken, and the usual bulk phases are stable. However, as is often the case in classical systems \cite{Chaikin1995}, we find that several other boundary conditions lead to a formation of a domain wall in the system. It is shown that such a domain wall induces an algebraic behavior in the off-diagonal elements of the single particle density matrix leading to an increase of the entanglement entropy. We find that the local effect of the domain wall on the ground state properties is rather minor, while we expect that the dynamical properties are more prominently affected \cite{DallaTorre2013}.

The 1D EBHM has been studied in many contexts. The ground state phase diagram has been accurately mapped out by DMRG in Refs.~\onlinecite{Rossini2012,Ejima2014PRL}, and also by quantum Monte Carlo~\cite{Batrouni2013, Batrouni2014, Batrouni2015}. Excitation spectrum and linear responses were studied in Refs.~\onlinecite{Kollath2010, DallaTorre2013, Gremaud2016}. Refs.~\onlinecite{Deng2011, Ejima2014PRA} give a detailed analysis of entanglement entropies. A dependence of the bulk properties on the boundary conditions was considered in Ref.~\onlinecite{Kurdestany2014}. We revisit the variants of open boundary conditions considered therein and propose a simple effective picture to rationalize the numerical results. In particular, we argue that the entanglement entropy of a system with a domain wall does not diverge as the system size increases, and that it is not an indication of a bulk superfluid phase.

The remainder of the paper is organized as follows. In Sec.~\ref{sec:mod}, the model and observables of interest are introduced. Moreover, we discuss the relation to antiferromagnetic spin-$1$ chains and boundary conditions that we use. Sec.~\ref{sec:corr} is devoted to correlation functions and order parameters. These results are then put into perspective by further calculations of entanglement entropies in Sec.~\ref{sec:entr}, followed by the conclusion in Sec.~\ref{sec:con}. Technical aspects that are not covered in the main text are discussed in the appendices.

\section{\label{sec:mod}Model}
The Hamiltonian of the 1D EBHM on an open chain is given by
\begin{multline}
	\hat{H}= -J \sum_{i=1}^{L-1} \left(\hat{a}_i^\dagger \hat{a}_{i+1} + h.c. \right)  \\
	+ \frac{U}{2} \sum_{i=1}^{L} \hat{n}_i(\hat{n}_i - 1) + V \sum_{i=1}^{L-1} \hat{n}_i \hat{n}_{i+1} , \label{eq:HEBHM}
\end{multline}
where $\hat{a}_i^\dagger$ and $\hat{a}_i$ are bosonic creation and annihilation operators for site $i$ and $\hat{n}_i$ denotes the corresponding number operator. The parameters $J$, $U$ and $V$ represent the strength of hopping, on-site and nearest-neighbour interaction, respectively. We focus on cases at or near unit filling factor ${\overline{n} = N/L \simeq 1}$ and set ${J=1}$ in the remainder of the paper.

\subsection{Observables}\label{sec:obs}
\begin{figure}[t]
	\centering
	\includegraphics[width=0.95\columnwidth]{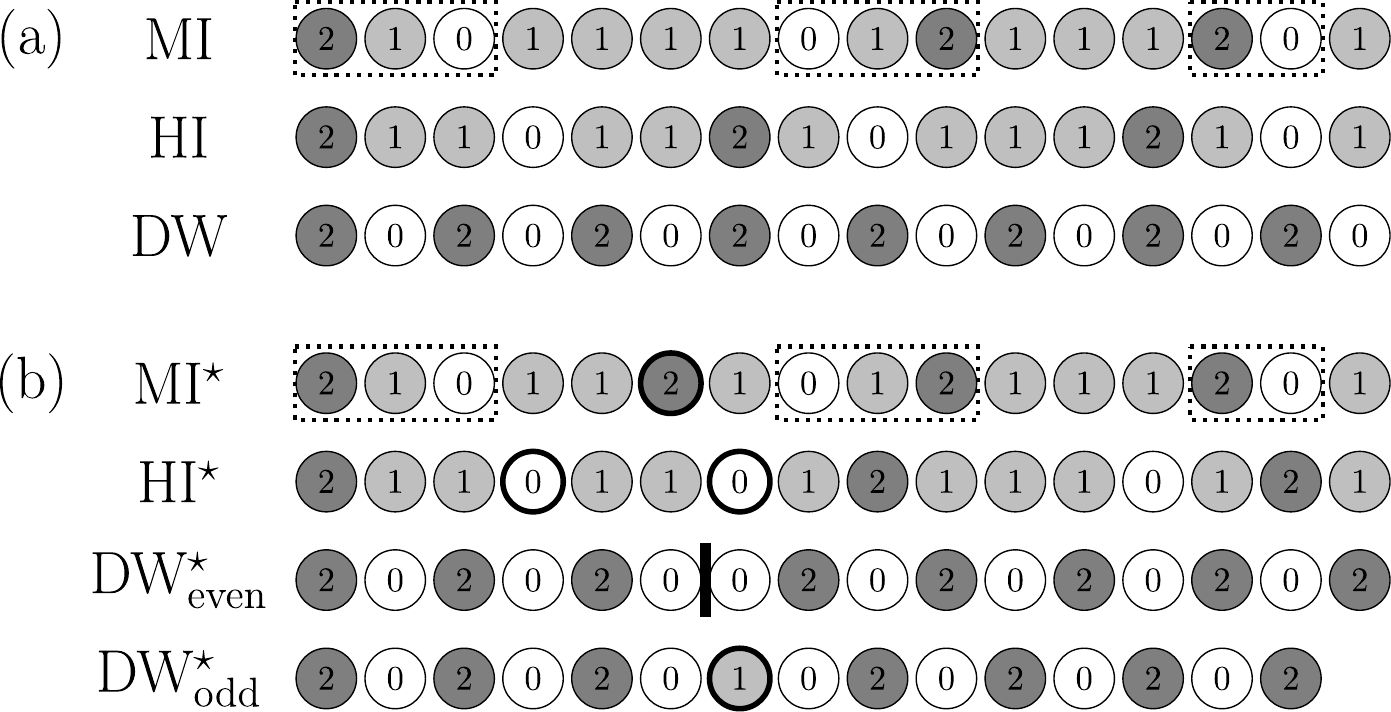}
	\caption{(a) Fluctuations pattern in a state with parity order (MI) or string order (HI), along with density wave occupation pattern without fluctuations (DW). The HI exhibits a dilute density wave order ${\ldots2\ldots 0\ldots 2\ldots 0\ldots}$, while the MI is characterized by bound doublon-holon pairs. (b) Single particle-like defects (MI$^\star$ and DW$_\text{odd}^\star$ [i.e. odd $L$]) and domain wall-like defects (HI$^\star$ and DW$_\text{even}^\star$ [i.e. even $L$]). In the HI and DW phases, the closest fluctuations to both edges are ${\delta n=1}$. For the DW phase with even $L$ and the HI, unit filling implies that there are two consecutive ${\delta n=-1}$ in the middle (thick circles).}
	\label{fig:patterns}
\end{figure}

The quantum phases of the 1D EBHM are characterized by the density wave, string, parity and superfluid correlation functions
\begin{equation}
	\begin{split}
	C_{\text{DW}}(i,j)&= \langle \delta\hat{n}_i(-1)^{i} \delta\hat{n}_j (-1)^{j}\rangle,  \\
	C_{\text{str}}(i,j)&= \langle \delta\hat{n}_i (-1)^{\sum_{k=i}^{j-1}\delta\hat{n}_k} \delta\hat{n}_j \rangle,  \\
	C_{\text{par}}(i,j)&= \langle (-1)^{\sum_{k=i+1}^{j-1}\delta\hat{n}_k} \rangle ,\\
	C_{\text{SF}}(i,j)&= \langle \hat{a}_i^\dagger \hat{a}_i \rangle , \label{eq:corr}
	\end{split}
\end{equation}
where ${\delta\hat{n}_i = \hat{n}_i-1}$. $C_\text{DW}$ measures the local staggered density modulations, while $C_\text{str}$ and $C_\text{par}$ are nonlocal. The latter two can be illustrated by considering the limit of large ${U\gg J}$, where the local Hilbert-spaces can be effectively reduced to occupation numbers ${n=0,1,2}$. A non-vanishing string order indicates a density wave pattern in the number fluctuations [see Fig.~\ref{fig:patterns}~(a)], where deviations from occupation number ${n=1}$ alternate between ${n=0}$ (holon) and $2$ (doublon), but appear at arbitrary distances~\cite{Berg2008}. Similarly, parity order indicates the presence of spatially bound doublon-holon pairs where doublons and holons do not necessarily alternate [see Fig.~\ref{fig:patterns}~(a)]. Finally, $C_\text{SF}$ measures the off-diagonal \mbox{(quasi-)} long range order of the single particle density matrix \cite{Yang1962}, indicating a gapless superfluid phase.

The order parameters $O_\text{DW}$, $O_\text{str}$ and $O_\text{par}$ are defined as the corresponding long distance limits of correlation functions~\cite{DallaTorre2006}
\begin{equation}
	{O=\lim_{|i-j|\to\infty}C(i,j)} .
\end{equation} 
One can identify the three gapped phases by these order parameters (see Table~\ref{tab:summary}): MI by ${O_\text{par}\neq 0}$ and ${O_\text{str}=O_\text{DW}=0}$, the HI by ${O_\text{str}\neq 0}$ and ${O_\text{par}=O_\text{DW}=0}$, and the DW phase by ${O_\text{str}\neq0\neq O_\text{DW}}$ (here $C_\text{par}$ oscillates). 

On the other hand, off-diagonal long range order does not exist for the one-dimensional case, even with arbitrarily weak interactions. Instead, SF order is marked by an algebraic decay ${C_\text{SF}(i,j)\sim |i-j|^{-K}}$ \cite{Lewenstein, XGWen}. Moreover, contrary to the gapped phases (MI, HI, and DW), the superfluid phase has gapless excitations. In one dimension, it behaves as a Tomonaga-Luttinger liquid, which is similar to critical modes near phase boundaries; both of them can be described by a conformal field theory. Such a critical behavior is reflected in the entanglement entropy. For any subsystem, the entanglement entropy is defined as the von Neumann-entropy of the corresponding reduced density matrix. For a subsystem of size $l$ that is the left block of a system of size $L$, we denote the entropy as $S_L(l)$. In this case, in a critical system, the entanglement entropy shows the following dependence on system and block size~\cite{Calabrese2004}
\begin{equation}
	S_L(l) = \frac{c}{6}\ln\left[\frac{2L}{\pi}\sin\Big(\frac{\pi l}{L}\Big)\right], \label{eq:entr}
\end{equation}
where $c$ is the central charge of the underlying conformal field theory. In case of a Tomonaga-Luttinger liquid, ${c=1}$ is expected. We will refer to the above formula to check whether certain boundary conditions can induce a SF phase. 

\begin{table*}[t!]
\caption{Nonzero DW, string and parity order parameters depending on $V$ and for each boundary condition. The on-site interaction is set to $U=6$ throughout. The superscript $\star$ indicates a sign-flip in the corresponding correlation function, an illustration of which is depicted in Fig.~\ref{fig:patterns}.}
\begin{ruledtabular}
	\begin{tabular}{cccc}
		& MI ($V \lesssim 3.5$) &  HI ($3.5 \lesssim V \lesssim 3.9$)  &  DW ($V \gtrsim 3.9$) \\
		\hline
		(A-even) &  parity &  string$^{\star}$ & DW$^{\star}$ + string$^{\star}$\\
		(A-odd) &  parity &  string$^{\star}$ & DW + string$^{\star}$ \\
		(B) &  parity &  string & DW + string \\
		(C-even) & parity$^{\star}$  $(V\lesssim 2.4)$; parity $(V \gtrsim 2.4)$
		&  string & DW$^{\star}$ + string$^{\star}$
	\end{tabular}
\end{ruledtabular}
\label{tab:summary}
\end{table*}

\subsection{Relation to spin-$1$ model}
As we have discussed in the introduction, the HI is an analog of the Haldane phase of antiferromagnetic spin-$1$ chains. Therefore, much of the common intuition about the strong coupling regime of the 1D EBHM stems from an effective spin-$1$ model introduced by Dalla Torre et al.~\cite{DallaTorre2006}. The effective spin model is obtained by restricting number fluctuations to ${|\delta \hat{n}_i| \leq 1}$ in the EBHM. By replacing ${\delta\hat{n}_i \to S_i^z}$ and ${\hat{a}_i^{(\dagger)} \to \hat{S}_i^{-(+)}}$, we find
\begin{equation}
	\hat{H}_\text{spin} = \sum_{i} \left(\hat{S}_i^x\hat{S}_{i+1}^x+\hat{S}_i^y\hat{S}_{i+1}^y + \Delta \hat{S}_i^z\hat{S}_{i+1}^z \right) + D \sum_{i} (\hat{S}_i^z)^2 , \label{eq:HeffSpin}
\end{equation}
where we have defined ${\Delta = V/2J}$ and ${D = U/2J}$. Setting the total magnetization to ${\sum_i\langle\hat{S}_i^z\rangle=0}$ is equivalent to unit filling in the 1D EBHM. For ${D,\Delta\gg 1}$, there exist three phases analogous to those of the 1D EBHM. These are defined by order parameters obtained from Eq.~\eqref{eq:corr} by substituting ${\hat{S}_i^z}$ for ${\delta\hat{n}_i}$ (we do not consider the components $\hat{S}^x$ and $\hat{S}^y$ here). The MI corresponds to a large-$D$ phase in the limit ${D\gg \Delta}$, the DW phase to a N\'eel phase in the opposite limit ${\Delta\gg D}$, and the HI to the Haldane phase in the intermediate regime where ${D\sim\Delta}$~\cite{Mikeska2005}. 

The ground states of the Haldane and N\'eel phases in an open chain show broken symmetries. More precisely, Kennedy and Tasaki showed that a so-called hidden ${\mathbb{Z}_2\times\mathbb{Z}_2}$ symmetry is broken in the Haldane phase~\cite{KennedyTasaki}. According to the valence-bond picture introduced by Affleck et al.~\cite{AKLT}, this corresponds to four degenerate configurations of effective spin-$1/2$ degrees of freedom at the two edges. The N\'eel phase holds a broken $\mathbb{Z}_2$ symmetry of spin flips ${\hat{S}_i^z \leftrightarrow -\hat{S}_i^z}$; in case of the open chain with even $L$, this is equivalent to broken lattice inversion symmetry.

Naively thinking, the ground states of corresponding phases in the 1D EBHM show similar degeneracies. For example, one might expect a broken lattice inversion symmetry in the DW phase (a particle-hole symmetry corresponding to the spin flips is obviously not present). We note, however, that the open boundary condition prohibits such degeneracy. This is because the nearest-neighbor term of the 1D EBHM requires additional local magnetic fields of strength $\Delta$ at the edges in the effective model,
\begin{equation}
	\begin{split}
	\sum_{i=1}^{L-1} \hat{n}_i \hat{n}_{i+1}
	&= \sum_{i=1}^{L-1} \left( \delta\hat{n}_i \delta\hat{n}_{i+1} + \hat{n}_i + \hat{n}_{i+1} \right) \\
	&\rightarrow \sum_{i=1}^{L-1} \hat{S}_i^z \hat{S}_{i+1}^z - \hat{S}_1^z - \hat{S}_{L}^z + \text{const.}.
	\end{split}
	\label{eq:edge_mag}
\end{equation}
Therefore, both the hidden ${\mathbb{Z}_2\times\mathbb{Z}_2}$ symmetry and spin-flip symmetry are not present in the effective Hamiltonian, while lattice inversion symmetry survives. We will see that the edge fields of Eq.~\eqref{eq:edge_mag} have the effect of pinning the edge spins, preventing a breaking of the remaining inversion symmetry. 

\subsection{Boundary conditions}\label{sec:BC}
As our discussion in the previous subsection indicates, a specific choice of boundary conditions can influence the ground state significantly. It is thus important to know how the bulk phases depend on the local boundary conditions. Experimentally, different boundary conditions might be chosen as an extra tuning knob to control quantum states.

In the present work we study an open chain of Eq.~\eqref{eq:HEBHM} with the following boundary conditions and filling factors~\cite{Kurdestany2014, Deng2011}:
\begin{itemize}
	\item[(A)] ${\overline{n}=1}$ and no further conditions,
	\item[(B)] ${\overline{n}=1}$ and opposite chemical potentials at the left and right edge sites: ${\mu(\hat{n}_1-\hat{n}_L)}$ with ${\mu=2J}$,
	\item[(C)] one extra boson ${N=L+1}$ (which implies ${\overline{n}\to 1}$ for ${L\to\infty}$) and no further conditions.
\end{itemize} 
The open boundary condition (A) in the 1D EBHM corresponds to extra magnetic fields at the edges in the spin-1 model as shown in Eq.~\eqref{eq:edge_mag}. Originally, (B) and (C) were meant to lift ground state degeneracies in the spin-1 model. First, we note that the edge potentials of (B) are substantially larger than what would be needed to numerically chose one of several degenerate ground states. The reason for this will become clear below. Second, the idea behind the extra boson in (C) is that it localizes close to the edges in the HI phase, contributing an effective magnetization of ${m=1/2}$ on each side~\cite{Deng2011}. With weak nearest neighbor interaction, however, such an extra particle may not be localized at the edges, and form a domain wall-like excitation in the middle of the chain (see Sec.~\ref{sec:corr}).

In the following sections, we closely inspect the previous assumptions and results by more extensive numerical calculations. Table~\ref{tab:summary} summarizes our main results. In several cases, the correlation functions flip their signs around the middle of the chain (see Sec.~\ref{sec:StrParDW}). This allows to define an order parameter with reversed sign in the thermodynamic limit. We also find that in some of the cases with sign-flipping order parameter superfluid correlations show a quasi-algebraic behavior (see Sec.~\ref{sec:superfluid}).

\section{Correlation functions}\label{sec:corr}
In this section we investigate string, density, parity and superfluid correlations in the ground states of the MI, HI and DW phases. To this end, we carry out MPS based DMRG calculations with site occupations truncated at ${n_\text{max}=4}$ and a maximum bond-dimension of ${\chi_\text{max}=250}$~\cite{openMPS}, which is sufficient for numerical convergence. Throughout the paper, we assume a large on-site interaction ${U=6}$, and most of the results are for up to ${L=250}$ sites. 

We first discuss the dependence of the three correlation functions, $C_{\text{DW}}$, $C_{\text{parity}}$, and $C_{\text{str}}$, and of the site occupation on the boundary conditions. Second, we show the dependence of the corresponding order parameters on $V$. Finally, we study the SF correlation functions. 

\subsection{String, density and parity correlations}\label{sec:StrParDW}
\subsubsection{Case (A)}
\begin{figure*}[t]
	\centering
	\includegraphics[width=\textwidth]{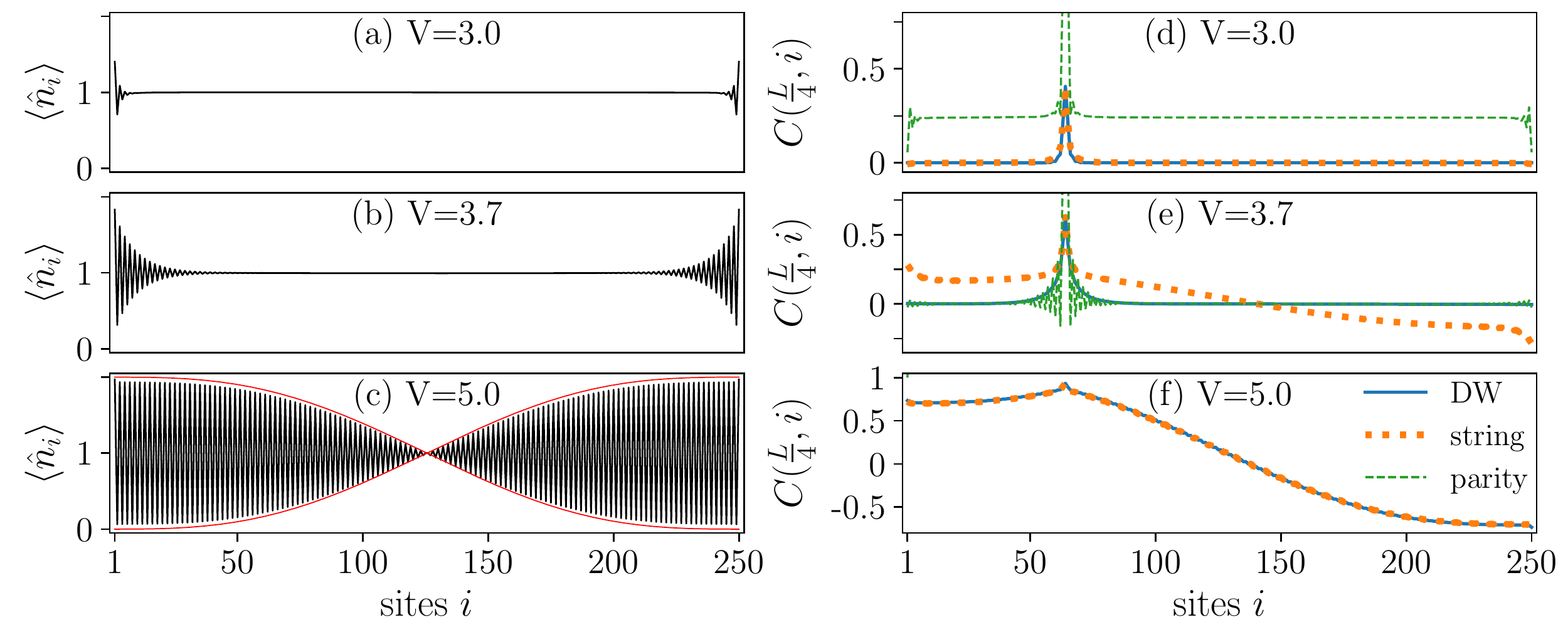}
	\caption{Boundary condition (A) for ${L=250}$, ${U=6}$ and from top to bottom: ${V=3.0}$ (MI), ${3.6}$ (HI), ${5.0}$ (DW). The left panels (a)-(c) depict occupation patterns. In (c) the red line represents the envelope calculated from the domain wall states in Eq.~\eqref{eq:pattern}. The right panels (d)-(f) show string (dotted orange), DW (solid blue) and parity (dashed green) correlation functions ${C(L/4, i)}$.}
	\label{fig:Ae}
\end{figure*}

\begin{figure*}[t]
	\centering
	\includegraphics[width=\textwidth]{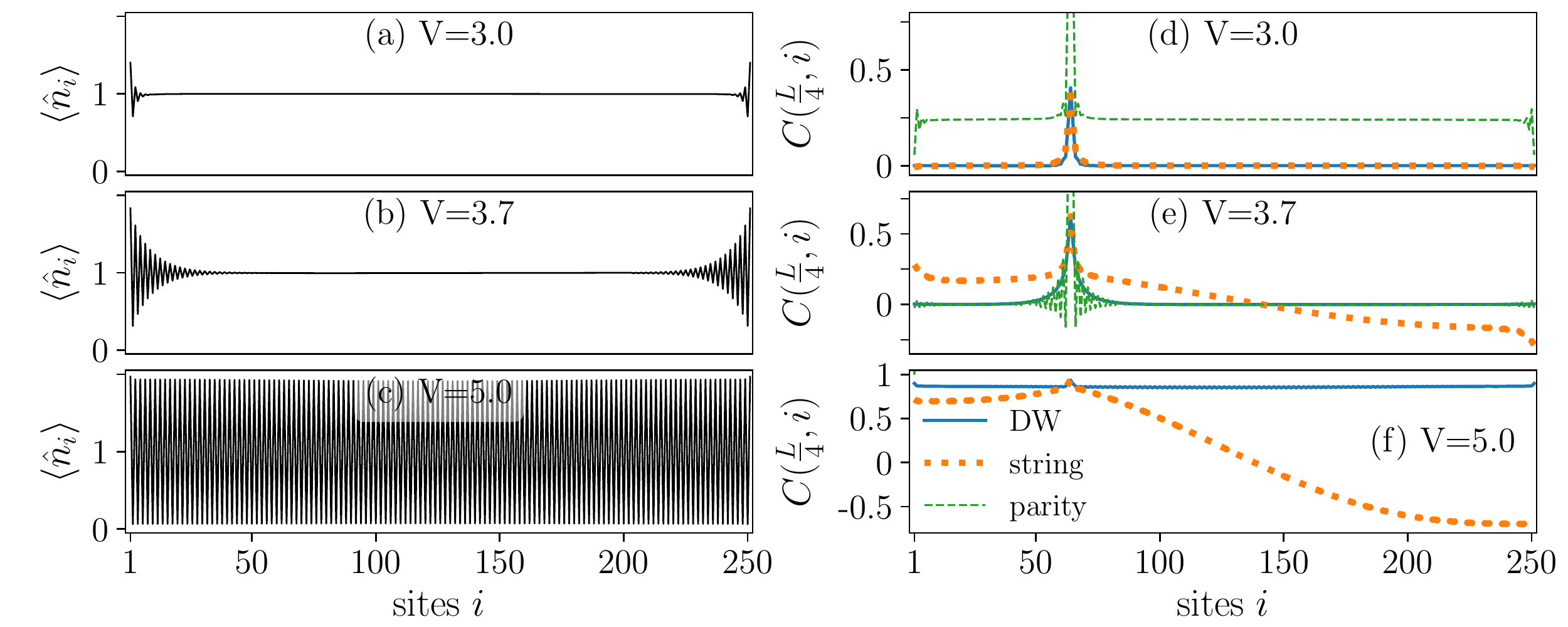}
	\caption{Boundary condition (A) for ${L=251}$, ${U=6}$ and from top to bottom: ${V=3.0}$ (MI), ${3.6}$ (HI), ${5.0}$ (DW). The left panels (a)-(c) depict occupation patterns, the right panels (d)-(f) show string (dotted orange), DW (solid blue) and parity (dashed green) correlation functions ${C(L/4, i)}$.}
	\label{fig:Ao}
\end{figure*}

Let us first consider boundary condition (A). Figs.~\ref{fig:Ae} and \ref{fig:Ao} show the site occupations and correlation functions for ${L=250}$ (A-even) and ${L=251}$ (A-odd), respectively. The left panels show the site occupations of selected ground states. In the DW phase of the even chain [see Fig.~\ref{fig:Ae}(c)], there is a change from an up-down to a down-up pattern in the middle of the chain. We checked that this pattern scales with the size of the system, which is why we call it a macroscopic boundary effect. In contrast, such a pattern does not appear in the odd case [see Fig.~\ref{fig:Ao}(c)]. 

In the right panels of Figs.~\ref{fig:Ae} and \ref{fig:Ao}, we depict the corresponding string, density and parity correlation functions evaluated between site ${\sim L/4}$ and all other sites $i$,
\begin{equation}
	C(L/4, i).
\end{equation}
One can clearly distinguish MI, HI and DW, depending on which correlations decay to zero; a more detailed discussion will be given in Sec.~\ref{sec:op}. We find that some of the correlation functions remain nonzero, while they flip signs in the middle of the chain. In the case of (A-even), the string correlation changes its sign in the HI phase, and both string and density correlations flip their signs in the DW phase [Figs.~\ref{fig:Ae}(e)-(f)]. On the other hand, in the case (A-odd) only the string correlation behaves in this way in the HI and DW phases [Figs.~\ref{fig:Ao}(e)-(f)]. The absence of a sign-flip of the density correlation function of the odd chain is consistent with the uniform occupation pattern in Fig.~\ref{fig:Ao}(c). 

The DW patterns can be understood by the fact that both edge populations are pinned to ${\langle\hat{n}_{1,L}\rangle \approx 2}$, which creates a domain wall in between. The edge populations are pinned because of the lack of one nearest-neighbor at the edge. In the spin model, this is encoded by the effective edge fields in Eq.~\eqref{eq:edge_mag}. Assuming a simple effective picture in the limit ${V\gg U\gg J}$, the dominant configurations for (A-even/odd) have a domain wall as
\begin{equation}
	\ket{\psi_j} =
	\begin{cases}
	\ket{20\ldots 2020}\ket{0202\ldots 02}  & \text{(A-even)}, \\
	\ket{20\ldots 2020}\ket{1}\ket{0202\ldots 02}   & \text{(A-odd)},
	\end{cases}
	\label{eq:pattern}
\end{equation}
where the domain wall is located at site $\sim 2j$ (see also Fig.~\ref{fig:patterns}~(b), cases DW$_\text{even/odd}^\star$). From this, it is clear that the staggered density modulations change signs only in the even case. As is shown in Appendix~\ref{app:domain}, the ground state approximately takes the following form,
\begin{equation}
	\ket{G} = \sum_j w_j \ket{\psi_j} \sim \sum_j \sin\left(\frac{\pi j}{d+1}\right)  \ket{\psi_j},
	\label{eq:dw_wavefunc}
\end{equation}
where $d$ is the number of domain wall states (here $d=L/2+1$) and ${j=1,\ldots,d}$ are labels of possible positions of the domain wall. This leads to the envelope indicated in Fig.~\ref{fig:Ae}(c) (see the red line). 

In contrast to the density correlation function, the string correlation function is not sensitive to even and odd distances. In the DW phase, the domain wall configurations in Eq.~\eqref{eq:pattern} both give spin-flipping string correlations. We can interpret the sign-flip of $C_\text{str}$ in the HI phase along the same lines. Because with boundary condition (A) doublons prefer to locate near the edges, a domain wall-like structure is created in between (see Fig.~\ref{fig:patterns}~(b), case HI$^\star$, for an illustration). This occurs regardless of whether $L$ is even or odd, and it is consistent with a sign-flip of string correlations. A toy wave function that captures this sign-flip of string correlations is given by \cite{Berg2008}
\begin{equation}
	\ket{\psi_j} = \hat{a}_1^\dagger \prod_{i=1}^{j-1} (\hat{a}_i^\dagger + \hat{a}_{i+1}^\dagger) \prod_{i=j+1}^{L-1} (\hat{a}_i^\dagger + \hat{a}_{i+1}^\dagger) \hat{a}_L^\dagger \ket{0}.
	\label{eq:HI_dw}
\end{equation}
The expectation value of a local density fluctuation $\braket{\delta n_i}$ of this state adds up to $\sim 1/2$ near the edges (i.e., edge states), and to $\sim - 1/2$ near the sites $j$ and $j+1$ corresponding to a domain wall. As will be discussed in Secs.~\ref{sec:superfluid}~and~\ref{sec:entr}, the superposition of the domain wall states $ \ket{G} \sim \sum_{j=1}^{L-1} w_j \ket{\psi_j}$ with $w_j$ given in Eq.~\eqref{eq:dw_wavefunc} leads to SF correlations and an entanglement entropy that qualitatively agree with the DMRG results.

\begin{figure*}[t]
	\centering
	\includegraphics[width=\textwidth]{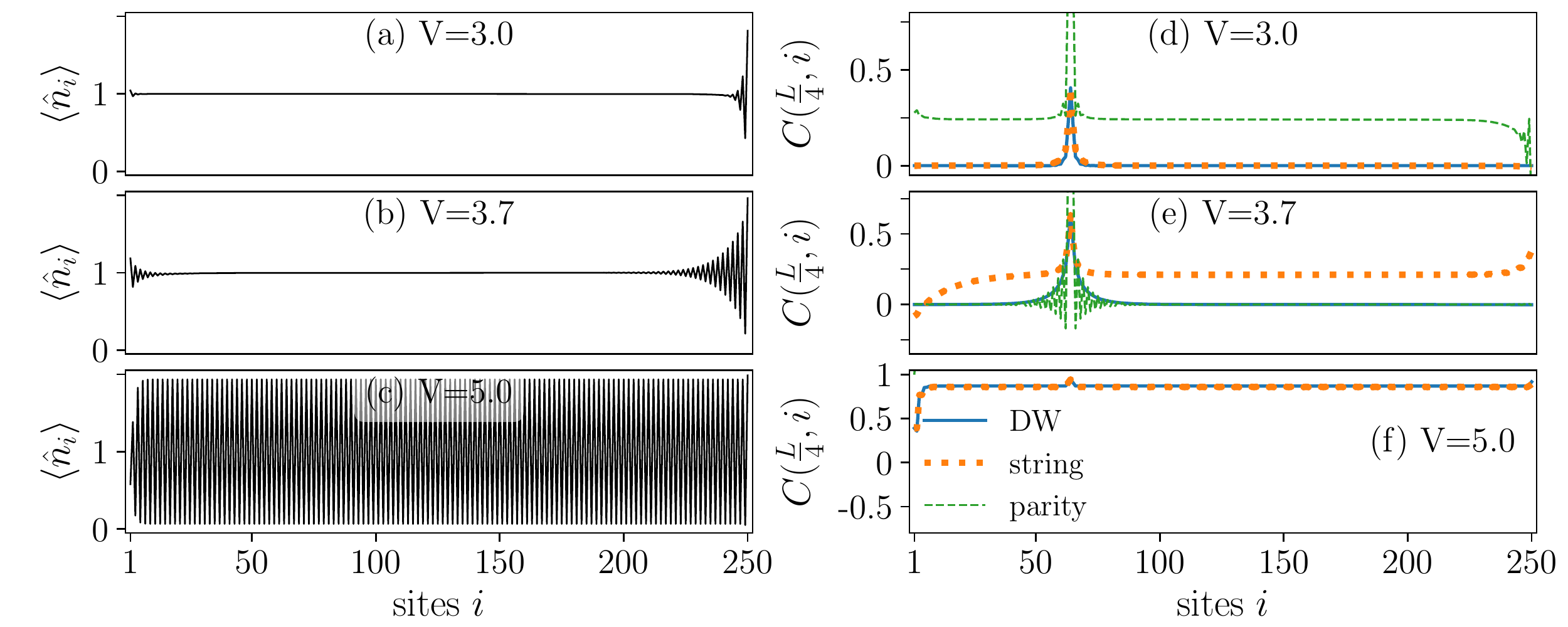}
	\caption{Boundary condition (B) for ${L=250}$, ${U=6}$ and from top to bottom: ${V=3.0}$ (MI), ${3.6}$ (HI), ${5.0}$ (DW). The left panels (a)-(c) depict occupation patterns, the right panels (d)-(f) show string (dotted orange), DW (solid blue) and parity (dashed green) correlation functions ${C(L/4, i)}$}
	\label{fig:B}
\end{figure*}
\begin{figure*}[t]
	\centering
	\includegraphics[width=\textwidth]{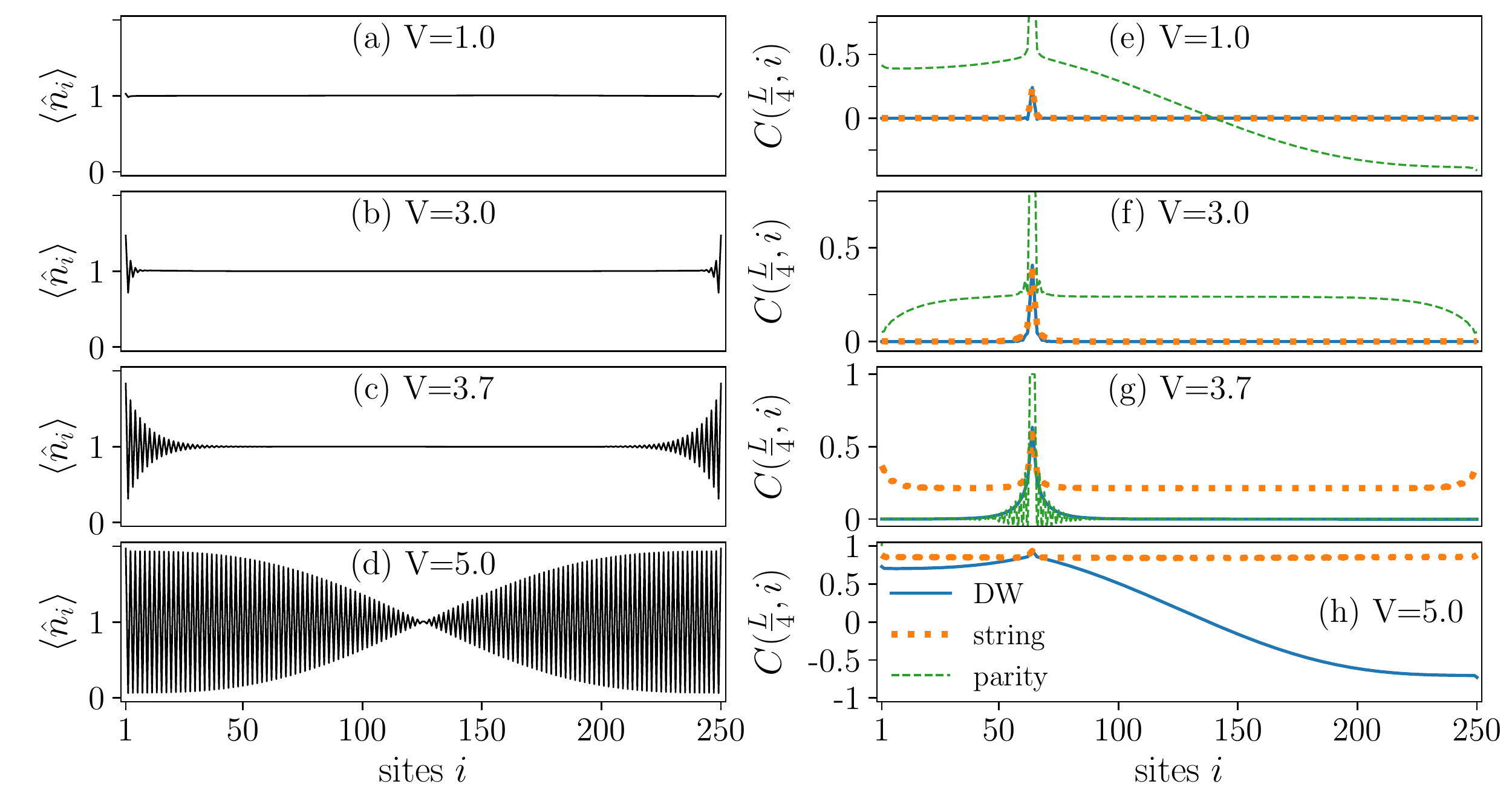}
	\caption{Boundary condition (C) for ${L=250}$, ${U=6}$ and from top to bottom: ${V=2.0,3.0}$ (MI), ${3.6}$ (HI), ${5.0}$ (DW). The left panels (a)-(d) depict occupation patterns, the right panels (e)-(h) show string (dotted orange), DW (solid blue) and parity (dashed green) correlation functions ${C(L/4, i)}$}
	\label{fig:C}
\end{figure*}

\begin{figure*}[t!]
	\centering
	\includegraphics[width=0.9\textwidth]{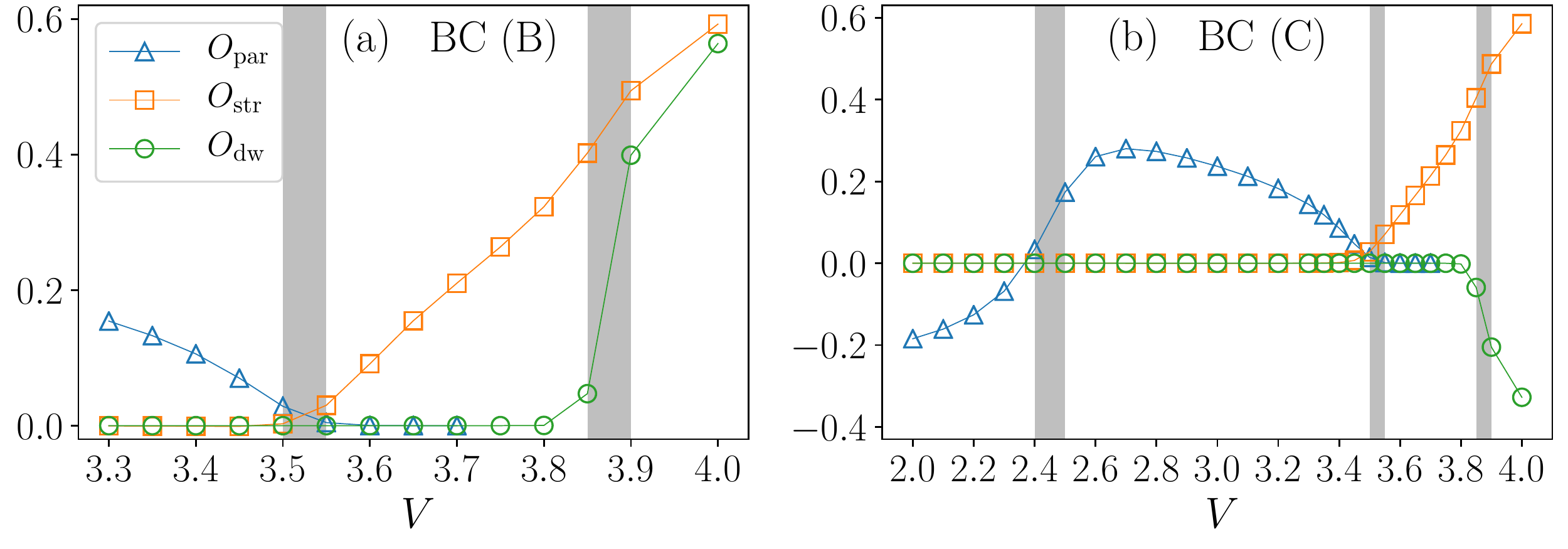}
	\caption{Order parameters $O_\text{par}$, $O_\text{str}$ and $O_\text{DW}$ as a function of $V$ for ${U=6}$ and ${L=250}$ with boundary condition (B) in panel (a) and boundary condition (C) in panel (b). Shaded areas indicate the position of critical points as narrowed down by a finite size analysis.}
	\label{fig:OPs}
\end{figure*}

\subsubsection{Case (B)}
Let us now turn to boundary condition (B), which is depicted in Fig.~\ref{fig:B}. In this case, the left edge potential, ${\mu >0}$, favors an empty state, while the right edge potential, ${-\mu <0}$, favors a doubly occupied state, and the inversion symmetry is broken. Therefore, the occupation pattern in the DW phase [Fig.~\ref{fig:B}(c)] as well as string and density correlations [Figs.~\ref{fig:B}(e)-(f)] show a uniform bulk behavior. The edge potentials of (B) have eliminated the macroscopic boundary effects of case (A). This is also seen in the entanglement entropies in Sec.~\ref{sec:entr} and in the entanglement spectra in Appendix~\ref{app:es}. Contrary to case (A), the edge chemical potential induces a slight asymmetry on the occupations around the left and right edges [Figs.~\ref{fig:B}(a)-(c)]. The large absolute value ${\mu=2J}$ of the edge potentials is necessary to remove the sign-flips of (A) in Fig.~\ref{fig:Ae}. This rules out the possibility of the sign-flips being a numerical artifact caused by ground state degeneracies. As we decrease the value of $\mu$, we find that the domain wall migrates from the edge to the middle of the chain. Only as the limit $V\to\infty$ is approached, where the domain wall states of Eq.~\eqref{eq:pattern} are indeed degenerate ground states, the value of $\mu$ necessary to remove the sign-flips goes to zero.

\subsubsection{Case (C)}
For condition (C), we have one extra boson, i.e. ${N=L+1}$. This extra boson has indeed restored a uniform string correlation in the HI and DW phase [Figs.~\ref{fig:C}(g)-(h)]. However, the density correlation function and the occupation pattern [Figs.~\ref{fig:C}(d)~and~(h)] in the DW phase show the same sign flip as for (A-even). This can again be illustrated in terms of a domain wall state. In the present case (C-even), the additional particle is placed on the domain wall state of (A-even), which can be effectively expressed by patterns such as
\begin{equation}
	\ket{20\ldots 203020\ldots 20}\ket{0202\ldots 02}.
	\label{eq:patternC}
\end{equation}
For these states, the sign flip disappears for $C_\text{str}$, but persists for $C_\text{DW}$. In Sec.~\ref{sec:entr} we argue based on the spatial dependence of the entanglement entropies that the extra boson is bound to the domain wall, and hence the relevant domain wall states in this case look like
\begin{equation}
	\ket{\psi_j} = \ket{20\ldots 20\underline{3}0}\ket{0202\ldots 02}.
	\label{eq:psi_C1}
\end{equation}

One difference of condition (C) compared to (A) and (B) appears in the small $V$ regime, ${V \lesssim 3.5}$. For the boundary conditions (A) and (B), this regime corresponds to the MI phase. In case (C), instead, we find two distinct regimes of nonzero parity order:	(i) for ${V \lesssim 2.4}$ the parity correlation function flips sign [see Fig.~\ref{fig:C}(e)] and (ii) for ${2.4 \lesssim V \lesssim 3.5}$ it settles to a constant value [see Fig.~\ref{fig:C}(f)]. The reason for this, as we show explicitly in Sec.~\ref{sec:occ}, is that the extra boson localizes at the edges when $V$ is increased. Therefore, in the regime (ii) with higher $V$, the bulk is effectively undoped. From the perspective of the doping, we then expect the lower $V$ regime to have nonzero SF correlations and the larger $V$ regime to be a pure MI, which we examine more carefully in Sec.~\ref{sec:superfluid}. However, the present results show that the state in the low $V$ regime must be distinct from a pure bulk SF state, because parity correlations decay to zero in the pure SF phase. We interpret it as a MI with additional off-diagonal long range correlations induced by states such as
\begin{equation}
	\ket{\psi_j} = \ket{1\ldots 1}\ket{2} \ket{1\ldots 1},
	\label{eq:121}
\end{equation}
which hosts an additional single particle at site $j$. Sufficiently far from the additional particle, fluctuations are still predominantly doublon-holon pairs (see Fig.~\ref{fig:patterns}~(b), case MI$^\star$).

\begin{figure*}[t]
	\centering
	\includegraphics[width=0.8\textwidth]{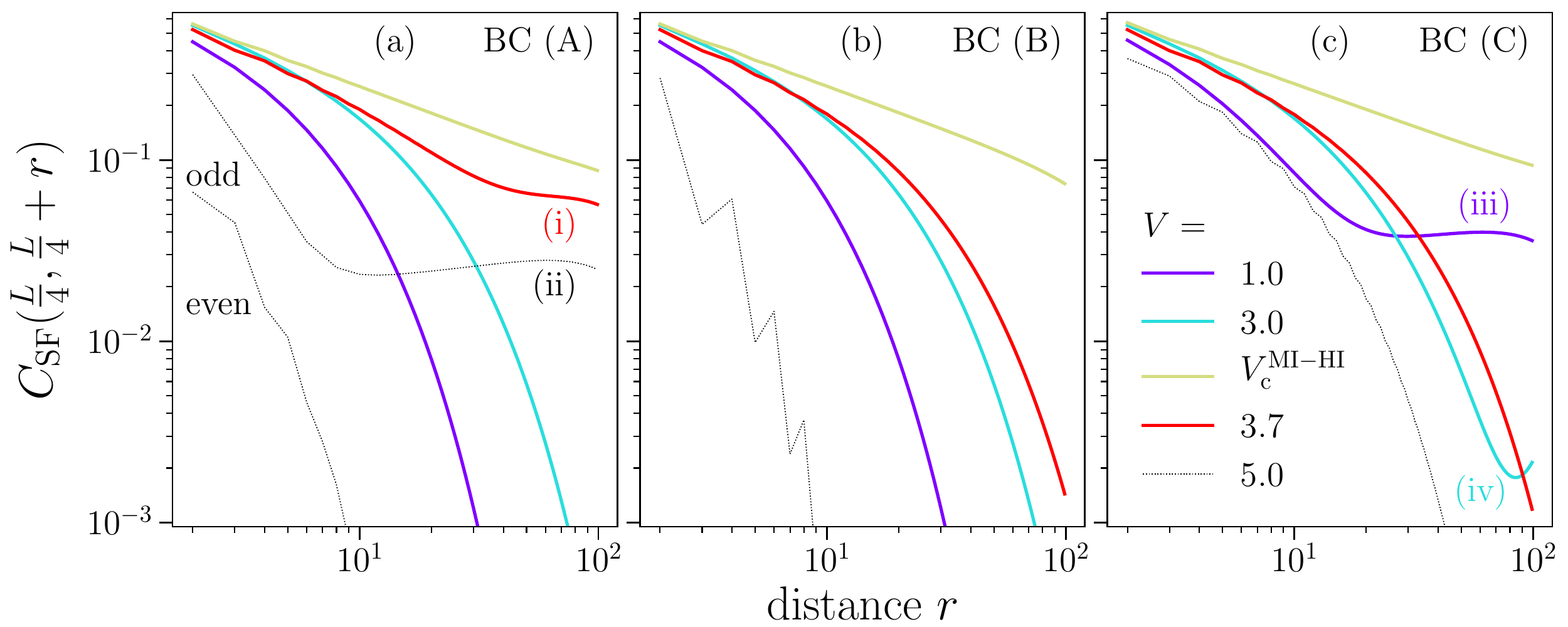}
	\caption{SF correlation functions ${C_\text{SF}(L/4,L/4+r)}$ for ${L=250}$ with boundary conditions (A), (B) and (C). The values of $V$ are chosen as ${V=1.0}$ [MI for (A) and (B), SF for (C)], ${V=3.0}$ (MI), ${V_\text{c}^\text{MI-HI}=3.5}$ for (A) and (B) and ${3.55}$ for (C) (closest to critical power law), and ${V=3.7}$ (HI). For the case of odd ${L=251}$ in the DW phase in panel (a), we only show values for even $r$ due to strong oscillations of $C_\text{SF}$. The colored labels (i)-(iv) mark the graphs that show significant boundary effects in the order that they are discussed in the text and shown in Fig.~\ref{fig:SF_scaling}.}
	\label{fig:SF_Vcomp}
\end{figure*}

\subsection{Order parameters}\label{sec:op}
Here we discuss the order parameters, $O_{\text{DW}}$, $O_{\text{parity}}$, and $O_{\text{str}}$. In practice, due to the finite size of the system, we choose ${O=C(L/4,3L/4)}$. Since this definition involves fixed sites relative to the system size, we can obtain well-defined values even for the sign-flipping cases.

First, we find that the positions of critical points do not significantly depend on the boundary condition. In Fig.~\ref{fig:OPs}, we show the order parameters as functions of $V$ for case (B), where edge potentials support well-defined bulk phases, and for case (C), where an extra particle, ${N=L+1}$, is added. In both cases, our finite size analysis narrowed down the MI-HI transition to ${V_\text{MI-HI}^c \approx 3.525 \pm 0.025}$ and the HI-DW transition to ${V_\text{HI-DW}^c \approx 3.875 \pm 0.025}$. For case (A-even/odd), the phase boundaries fall into the same range (not shown). The robustness of the phase boundaries to different boundary conditions is due to the fact that the influence of a domain wall on the local properties of the state vanishes in the thermodynamic limit. The negative value of $O_\text{DW}$ in case (C) is due to the sign-flip of $C_\text{DW}$.

Second, in Fig.~\ref{fig:OPs}(b) for case (C), one can see an extra transition from a sign-flipping to constant parity correlation function around ${V\approx 2.4}$, which does not exist for boundary conditions (A) and (B). As is shown below, this additional transition is accompanied by a transition from algebraic to exponential decay of SF correlations, which is caused by the localization of the extra particle to the edge sites due to strong $V$.

\subsection{Superfluid order}\label{sec:superfluid}

Let us now take a look at the spatial dependence of superfluid correlations. To this end we compare the boundary conditions (A), (B) and (C) in Fig.~\ref{fig:SF_Vcomp}. For each case, we plot ${C_\text{SF}(L/4,L/4+r)}$ with ${L=250}$ for four selected ground states far from criticality and one ground state close to the MI-HI transition. For boundary condition (A) and ${V=5}$, representing the DW phase, we have included the case of an odd chain, ${L=251}$, because it differs from the even chain.

\begin{figure*}
	\centering
	\includegraphics[width=\textwidth]{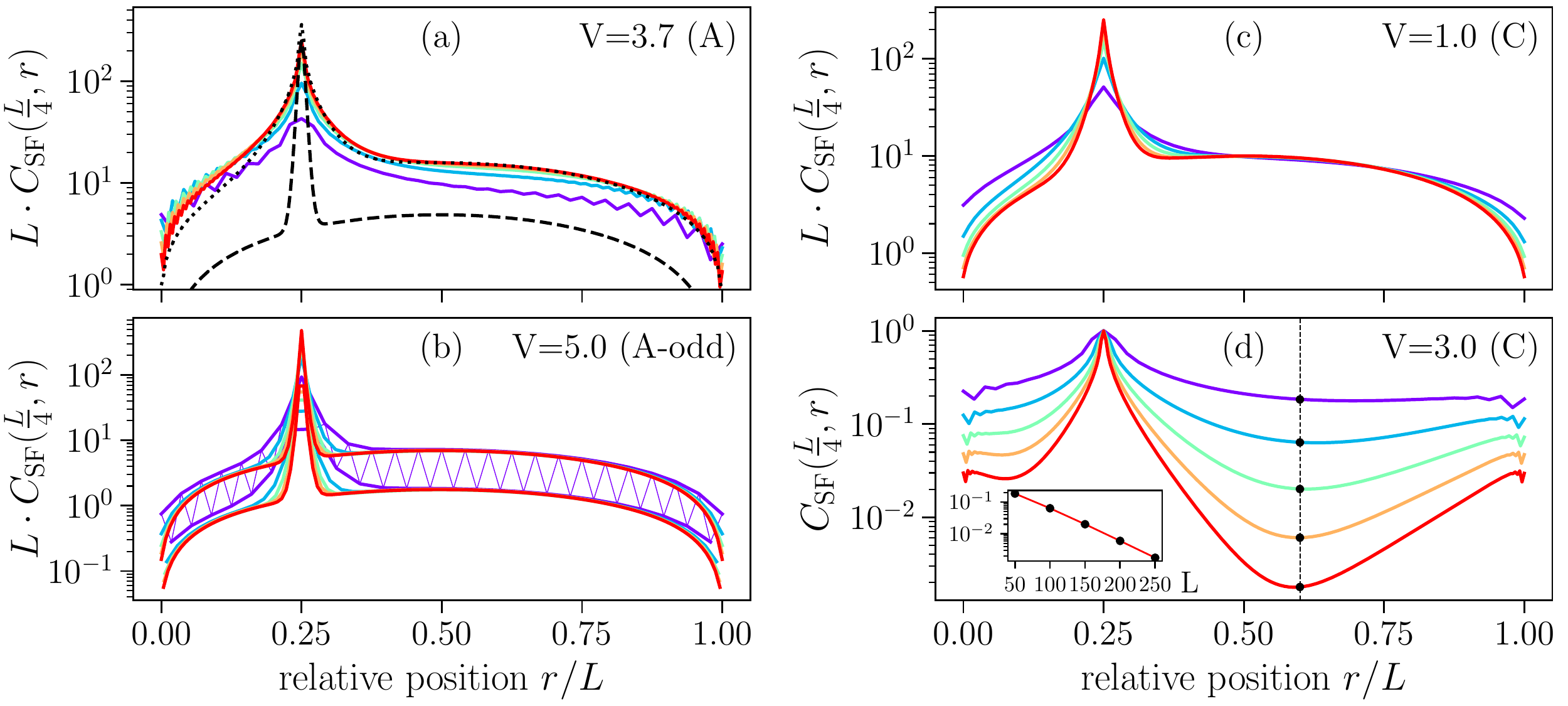}
	\caption{Finite size scaling of SF correlation functions [from ${L=50}$ (purple) to ${L=250}$ (red)] for the four cases of Fig.~\ref{fig:SF_Vcomp}, which are neither power law nor exponential. In panels (a), (b) and (c) we have rescaled $C_\text{SF}$ by the factor $L$. The dotted plot in (a) shows a DMRG calculation of the spin-$1$ model \eqref{eq:HeffSpin} with edge fields \eqref{eq:edge_mag} for ${U=5}$, ${V=3.4}$ (Haldane phase), where ${\langle\hat{a}_i^\dagger\hat{a}_j\rangle}$ has been replaced by ${\langle\hat{S}_i^+\hat{S}_j^-\rangle}$; the dashed line in (a) comes from a simple domain wall picture for the HI phase. In case (b) the correlations at even and odd distances are plotted separately and oscillations are shown for ${L=50}$. The inset of panel (d) indicates exponential decay of ${C_\text{SF}(0.25\cdot L, 0.6\cdot L)}$ as a function of $L$.}
	\label{fig:SF_scaling}
\end{figure*}

In all three cases, $C_\text{SF}$ decays by power-law near the MI-HI transition (yellow lines) due to quantum critical fluctuations. On the other hand, $C_\text{SF}$ decays exponentially in the MI phase (light blue lines, ${V=3.0}$) regardless of the boundary conditions. With boundary condition (C), $C_\text{SF}$ goes up again at large distances due to the extra particle localized close to the edges. With ${V=1}$ (purple lines), the boundary conditions (A) and (B) again give a MI phase and $C_\text{SF}$ decays exponentially. By contrast, the boundary condition (C) leads to a power law decay of $C_\text{SF}$ at short distances, followed by a plateau. Here the extra particle is delocalized over the whole chain, as indicated by the sign-flip of the parity correlation function in the previous section. In the HI phase (red lines, ${V=3.7}$), the SF correlations decay exponentially with boundary conditions (B) and (C). However, with boundary condition (A) in the sign-flipping HI phase, the correlations are enhanced. Here we see a power law decay followed by a plateau at larger distances. Finally, in the DW phase with ${V=5}$ (grey lines) we see exponential decay of $C_\text{SF}$ for (A-even), (B) and (C), whereas (A-odd) gives a power law followed by a plateau.	

We further investigate the system size-dependence of the four cases where $C_\text{SF}$ decays neither exponentially nor purely in a power-law manner (see labels in Fig.~\ref{fig:SF_Vcomp}): (i) ${V=3.7}$ with BC-(A), (ii) ${V=5.0}$ with BC-(A-odd), (iii) $V=1$ with BC-(C), and (iv) ${V=3.0}$ with BC-(C). In the first three cases, (i)-(iii), the influence of the edge appears as a macroscopic plateau, and we plot $C_\text{SF}$ rescaled by the factor $L$ in Figs.~\ref{fig:SF_scaling}(a)-(c),
\begin{equation}
	\tilde{C}_\text{SF} (r_1,r_2; L) \equiv L C_\text{SF} (r_1 L, r_2 L),
\end{equation}
where $r_{1,2} \in [0, 1]$ denotes the relative position in the chain. The convergence of the curves indicates that, as ${L\to\infty}$, the plateaus scale as
\begin{equation}
	C_\text{SF} (r_1 L, r_2 L) \sim L^{-1}.
\end{equation}
Therefore, we conclude that these states show quasi-long range superfluid correlations. The similarity among (i), (ii) and (iii) suggests that in all three cases a single-particle-like excitation is responsible for the quasi-long range order.  

On the other hand, for case (iv) [Fig.~\ref{fig:SF_scaling}(d)], we find that as a function of $L$
\begin{equation}
	\tilde{C}_\text{SF}(r_1, r_2; L) \sim L e^{- L \gamma(r_1,r_2)}
\end{equation}
for ${r_1=0.25}$ and $r_2 \in [0, 1]$. For roughly ${r_2 \in[0.3,0.5]}$ the factor in the exponent behaves as ${\gamma(r_1, r_2)\sim |r_1-r_2|}$, which corresponds to an undoped MI. If we plotted $\tilde{C}_\text{SF,L}$ for various $L$'s in absolute distance scales, they would lie on top of each other. On the other hand, when ${r_2 \gtrsim 0.6}$, the correlations increase again, due to the extra particle. However, the value at the edge, ${\tilde{C}_\text{SF}(0.25,1; L)}$, still decays exponentially as the system size $L$ increases, and hence we conclude that there is no superfluid order in this case.

For case (i), we can compare our DMRG results on $C_\text{SF}$ with an analytical estimate given by the domain wall wave function in Eq.~\eqref{eq:dw_wavefunc} with basis states from Eq.~\eqref{eq:HI_dw} [dashed black line in Fig.~\ref{fig:SF_scaling}(a)]. The latter leads to a qualitatively similar correlation function as the numerically exact one calculated by DMRG. However, with the toy wave function, $C_\text{SF}$ decays much faster at short distances and its absolute values are almost an order of magnitude lower.

Naively, the discrepancy may come from the lack of higher occupations $n>2$ in the toy wave function. To check this assumption, we have performed a DMRG calculation of the spin-$1$ model from Eq.~\eqref{eq:HeffSpin} with edge fields as in Eq.~\eqref{eq:edge_mag} for interaction parameters belonging to the Haldane phase [see the dotted black line in Fig.~\ref{fig:SF_scaling}(a)]. By analogy, $C_\text{SF}$ is replaced by ${\langle\hat{S}_i^+\hat{S}_j^-\rangle}$. Then the spin model exhibits the same short range behavior and order of magnitude of $C_\text{SF}$ as we see for the 1D EBHM. This shows that large local particle number fluctuations $n >2$ are irrelevant in this case in contrast to a bulk superfluid, and are not responsible for the discrepancy of the toy model and the DMRG results. Rather, we expect that the actual width of the domain wall is larger than the one of the toy wave function, since the kinetic energy term becomes more important for weaker interactions.

\subsection{Edge occupations}\label{sec:occ}

\begin{figure}[t]
	\centering
	\includegraphics[width=\columnwidth]{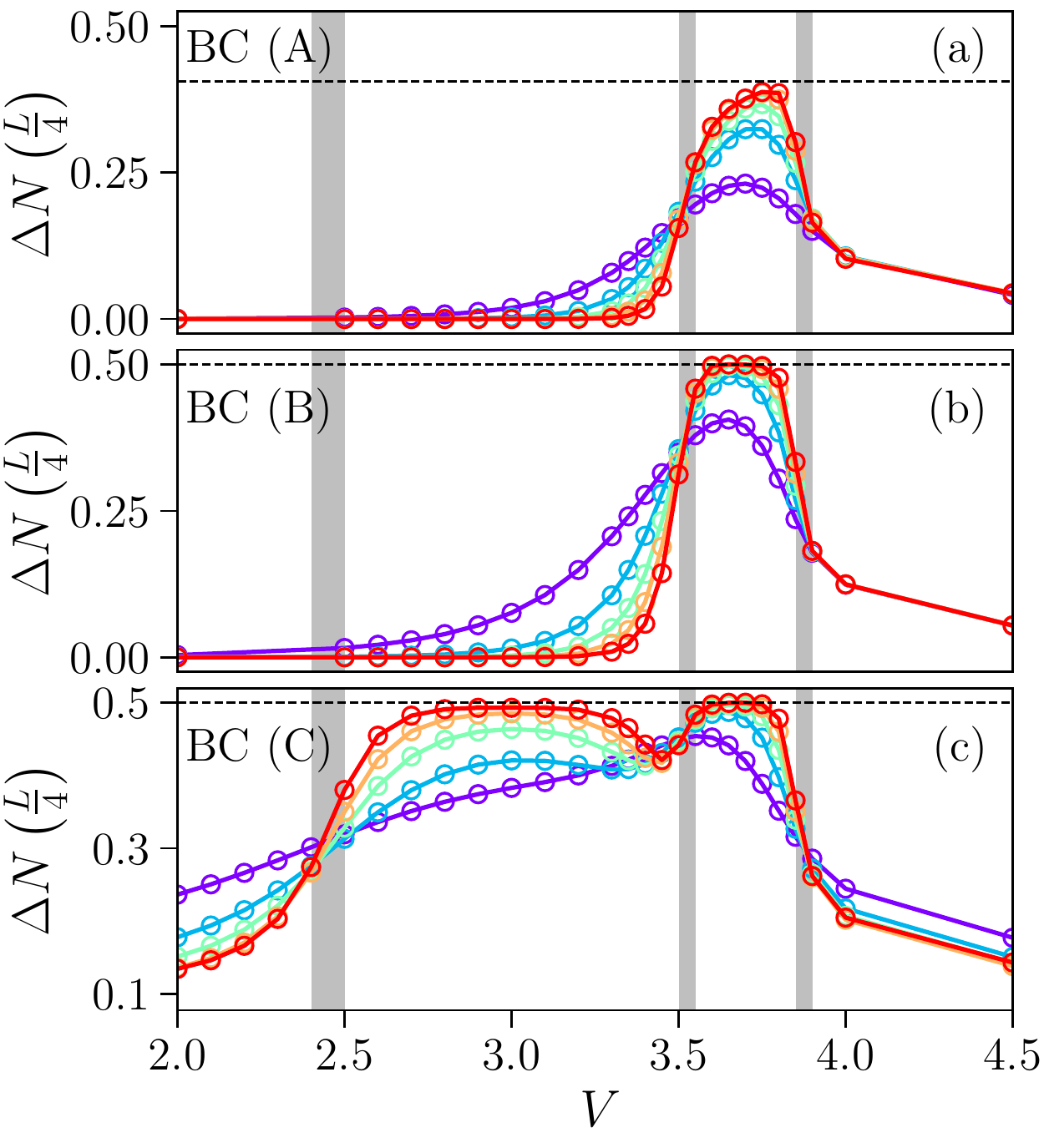}
	\caption{Excess densities ${\Delta N}$ of the quarter of the chain close to the right edge as a function of $V$ for system sizes from ${L=50}$ to ${L=250}$. Reading from top to bottom, the boundary conditions (A), (B) and (C) were used. In panel (a) the dashed black line indicates the value due of ${\Delta N}$ for a domain wall state as in Eq.~\eqref{eq:HI_dw} with a sine-like wave function. Panel (b) shows occupation numbers from the edge with negative potential.}
	\label{fig:n}
\end{figure}

One of the main arguments of the previous sections was based on an increased particle density close to the edges and hence a doping of the bulk. To confirm this interpretation, in Fig.~\ref{fig:n} we plot the excess population 
\begin{equation}
\Delta N(l) = \left( \sum_{i=1}^{l} \langle \hat{n}_i \rangle -1 \right)
\end{equation}
of the first ${l=L/4}$ sites as a function of $V$ for system sizes from ${L=50}$ to $250$.

For boundary condition (A) we do not see a quantization of ${\Delta N}$ to ${1/2}$ [Fig.~\ref{fig:n}(a)]. However, increased edge population in the HI phase, ${3.5\lesssim V\lesssim 3.9}$, is consistent with the value given by the toy states in Eq.~\eqref{eq:HI_dw} [dashed black line in panel~(a)]. On the other hand, the edge potentials of boundary condition (B) and the extra boson of (C) lead to $1/2$-edge states in the HI phase [Figs.~\ref{fig:n}(b)~and~(c)]. In addition, in case (C), one clearly sees that for $2.4\lesssim V\lesssim 3.5$ the extra particle localizes at the edges. This confirms our interpretation of the correlation functions and the plateau of the entanglement entropy as representing a MI phase.

\section{Entanglement entropy}	\label{sec:entr}

In Sec.~\ref{sec:corr} we found that certain boundary conditions induce quasi-long range SF order in parameter regimes which typically belong to the MI, HI or DW phase, where correlation functions of the original insulating phases flip merely their signs. In order to check, whether the quasi-long range SF correlations indeed imply the critical properties of a bulk SF phase or are merely an artifact of a single defect, we now investigate the entanglement entropies. If the system is truly in a bulk SF phase, as mentioned in Sec~\ref{sec:obs}, the block entropies ${S_L(l)}$ are marked by a characteristic dependence on the position $l$ of the bipartition [see Eq.~\eqref{eq:entr}] as well as a logarithmic divergence in the system size $L$ (volume law). On the other hand, if the SF correlations come from a domain wall, we expect a finite contribution to the entropy that does not diverge as $L$ increases. 

\begin{figure*}[t]
	\centering
	\includegraphics[width=\textwidth]{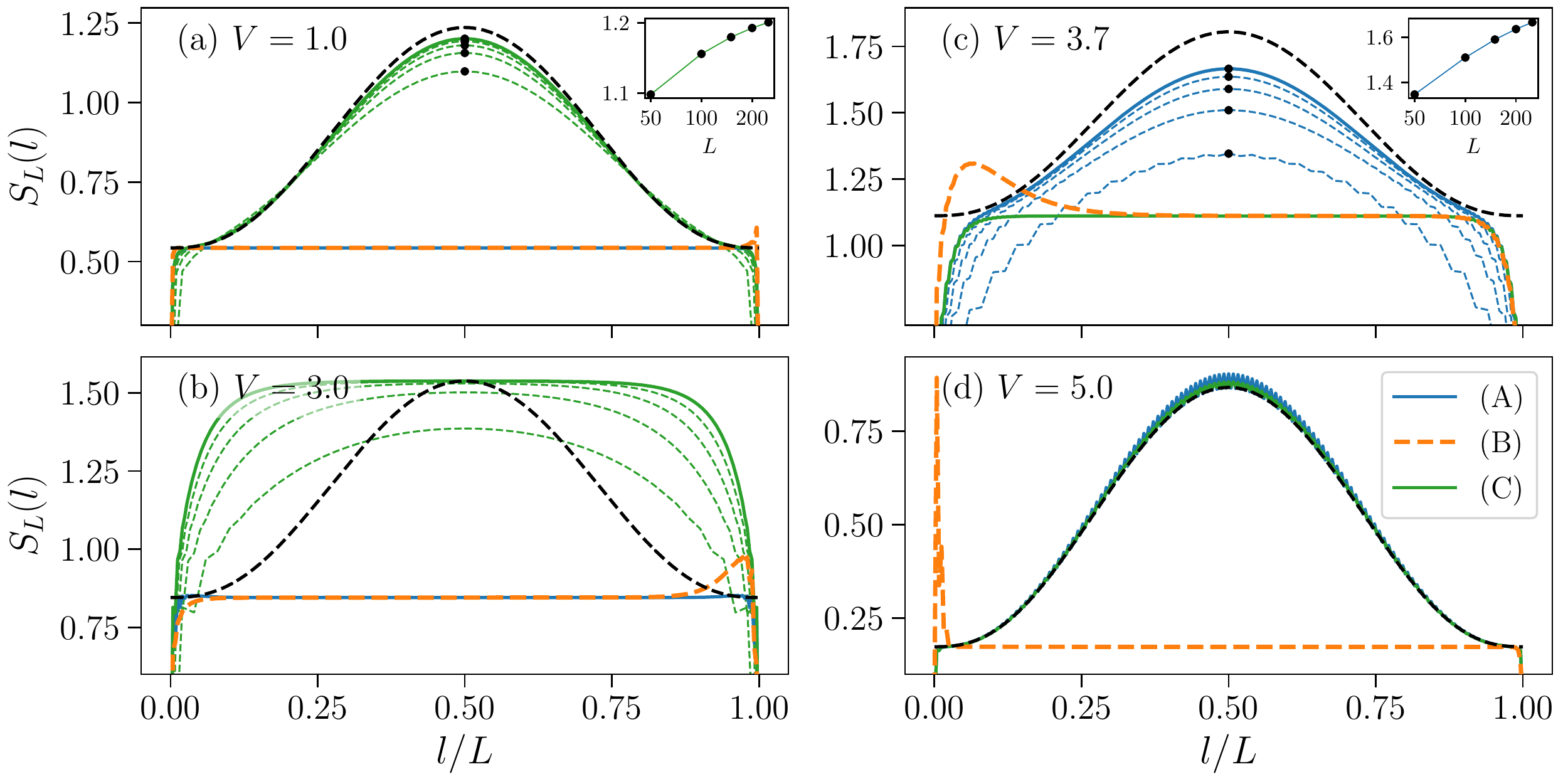}
	\caption{Block entropies of the left block as a function of block size over system size ${l/L}$ for ${L=250}$. Boundary conditions (A) [blue lines], (B) [green lines] and (C) [dashed orange lines] are compared for ${V=1.0,3.0,3.7,5.0}$ [panels (a), (b), (c) and (d), respectively]. Dashed black lines are the values calculated from the effective theory [see Appendix.~\ref{app:gen}] added onto the plateau value of boundary condition (B). Thin dashed lines in (a), (b) and (c) represent system sizes ${L=50,100,150,200}$ for (C) and (A), respectively. The insets show the values marked by black dots in the main figures, plotted against the logarithmic system size.}
	\label{fig:entr_FSS}
\end{figure*}

 In Fig.~\ref{fig:entr_FSS}, we plot the block and system size dependence of the entanglement entropies, focusing on the same representative cases ${V=1.0,3.0,3.7,5.0}$ as in Sec.~\ref{sec:corr}. In every case, we see a saturated plateau of ${S_L(l)}$ that does not depend on $L$ for boundary condition (B). This agrees with the absence of quasi-long-range SF correlations. If there is a single domain wall on top of a regular bulk background, we expect that the entanglement entropy can be written as a sum of two contributions. In order to calculate the contribution from the domain walls, let us consider a state $\ket{G}$ given by the superposition of local domain wall wave functions $\ket{\psi_j}$ e.g., given in Eqs.~\eqref{eq:pattern}, \eqref{eq:HI_dw}, \eqref{eq:psi_C1}, or  \eqref{eq:121}. We assume that the weight of each $\ket{\psi_j}$ is given by Eq.~\eqref{eq:dw_wavefunc}, i.e.
\begin{equation}
	\begin{split}
	\ket{G} &= \sum_{j = 1}^{d} w_j \ket{\psi_j}\\
	&= \sum_{j < m_l} \sin\left(\frac{\pi j}{d+1}\right) \ket{\psi_j} + \sum_{i\geq m_l} \sin\left(\frac{\pi j}{d+1}\right) \ket{\psi_j},
	\end{split}
	\label{eq:decomp}
\end{equation}
where the two terms correspond to the domain wall states whose centers are left or right of the bipartition point between sites $l$ and $l+1$, and $m_l$ is the smallest index of the domain wall in the right block. The two contributions in Eq.~\eqref{eq:decomp} can have an overlap due to configurations where the domain wall is close to site $l$, either on the left or the right, due to a finite width of the domain wall ($\braket{\psi_{j^\prime}|\psi_j}\neq 0$, see discussion below). However, in the thermodynamic limit the local influence of the domain wall at site $l$ vanishes. Hence, we ignore the overlap of the two contributions. In this limit, as we show in Appendix~\ref{sec:appB}, the entropy from the domain wall state $S^\text{extra}_L(l)$ simplifies to
\begin{equation}
	S^\text{extra}_L(l) \xrightarrow[l/L=\text{const.}]{L\to\infty} -p_l\ln(p_l)-(1-p_l)\ln(1-p_l)
	\label{eq:Slim},
\end{equation}
with
\begin{equation}
	p_l = \frac{\sum_{i\leq m_l}\sin^2\left(\frac{\pi i}{d+1}\right)}{\sum_{i\leq d}\sin^2\left(\frac{\pi i}{d+1}\right)}. \label{eq:pexplicit}
\end{equation}
We note that the result is equivalent to the entanglement entropy for the ground state of a tight-binding model with a single particle in an open chain. This naturally explains the similarly of the results for the domain walls and the extra particle.

For comparison with DMRG, we add the plateau value from boundary condition (B) to the domain wall contribution in Eq.~\eqref{eq:Slim}, and plot the result in Fig.~\ref{fig:entr_FSS} (dashed black lines).\footnote{In the DW phase we have not accounted for site-to-site oscillations of the probabilities $p_l$.} This is different from the result of a conformal field theory [see Eq.~\eqref{eq:entr}], which excludes the possible gapless phase that was anticipated in Ref.~\cite{Kurdestany2014}.

In the low $V$ regime, ${V\lesssim 3.5}$ [see Figs.~\ref{fig:entr_FSS}(a)~and~(b)], boundary condition (A) gives the same plateau as (B). With boundary condition (C), the entropy ${S_L(L/2)}$ of the symmetric bipartition saturates at ${\approx\ln 2}$ above this plateau; the inset of panel (a) shows that  ${S_L(L/2)}$ does not diverge logarithmically. For ${V=1}$ the dependence on the bipartition converges to the prediction of the sine-like wave function [see panel~(a)]. Therefore, the extra boson is approximately described by this wave function. Meanwhile, for ${V=3}$, we have a broad plateau, which is due the localization of the extra boson at the edges [see panel~(b)]. In both cases, finite size effects can be understood by additional number fluctuations, which are due to the extra boson. These cause finite overlaps of the effective basis states in Eq.\eqref{eq:decomp}.

In the HI regime, ${3.5\lesssim V\lesssim 3.9}$, boundary condition (C) agrees with the bulk value [see Fig.~\ref{fig:entr_FSS}(c)]. Here, the entropies for boundary condition (A) approach the domain wall prediction, and ${S_L(L/2)}$ increases slower than logarithmically [see the inset of panel~(c)]. However, the convergence is much slower than in the single particle case of panel (a). The slow convergence of the entropies points to a large overlap of the contributions of Eq.~\eqref{eq:decomp}. This overlap can be qualitatively understood in terms of Fig.~\ref{fig:patterns}~(b), case HI$^\star$. Namely, the domain wall can lie anywhere between the two occupations of $n=0$ that are highlighted in the figure, which indicates a larger width of the domain wall.

As for the DW phase, ${V=5.0}$, we see the domain wall picture very well confirmed by the DMRG simulations for both boundary condition (A) and (C) [see Fig.~\ref{fig:entr_FSS}(d)]. We recall the low energy effective states of boundary condition (C) [see Eq.~\eqref{eq:patternC}], where the extra boson is added to the effective states of boundary condition (A) [see Eq.~\eqref{eq:pattern}]. Naively, one might expect that this further increases the entropy up to an additional factor of ${\sim\ln 4}$ instead of ${\sim\ln 2}$ above the bulk value, because both the extra boson and the domain wall can either be on the left or on the right. The fact that the entropy is only increased by ${\sim\ln 2}$ can be understood by the extra boson being bound to the domain wall. Let us consider the configurations of Eq.~\eqref{eq:psi_C1}, where the extra boson is next to the domain wall. These are energetically favored, because virtual configurations ${\ldots 2021|0202\ldots}$ (next to the domain wall) are favored compared to ${\ldots 202120\ldots}$ (away from the domain wall) due to the nearest-neighbor interaction. 

Altogether we conclude that in all cases a single particle-like defect is responsible for enhanced SF correlations. Since the entanglement entropy does not diverge with the system size, these do not indicate a bulk SF phase. However, we expect that superfluid order would emerge if we fixed the doping density instead of doping with a single particle as in case (C).

\section{\label{sec:con}Conclusion}
In this work we investigated the influence of various open boundary conditions on the ground states of the 1D extended Bose-Hubbard model at or near unit filling. We found that the simple open chain at exactly unit filling has a non-degenerate ground state, even in the Haldane insulator and density wave regimes. In particular this means that the Haldane insulator does not have two degenerate edge state configurations, as would be naively expected. This can be explained by the fact that the nearest neighbor interaction $V$ effectively induces attractive edge potentials, causing a higher population of the edges and the presence of a domain wall around the middle of the chain. If the system is doped with a single extra boson, the regime corresponding to the Mott insulator splits in two regimes, where the extra boson behaves like the domain walls for low $V$ and is trapped at the edges for larger $V$.

In most cases the domain wall induces algebraic behavior of the off-diagonals of the single particle density matrix. However, this is not an indication of a fundamental change of the bulk order, but an artifact of the domain wall. Interestingly, in the density wave regime with even $L$, the domain wall has a different character, which leads to a modulation of the density wave pattern on the scale of $L$, but not to quasi-long range order.

Our results demonstrate that boundary conditions can have a significant influence on quantum many-body states, if interactions beyond contact-interaction are present. Local potentials at the edges may be used to control the effective filling factor. On the other hand, such effects can be a hindrance to the physical control of edge states, as these can be naturally pinned to one value.

\section{Acknowledgments}
The authors thank M. Thoss for helpful comments on the manuscript. We acknowledge support by the state of Baden-W\"urttemberg through bwHPC and the German Research Foundation (DFG) through Grant No. INST 40/467-1 FUGG. J.O. acknowledges support from Research Foundation for Opto-Science and Technology and from Georg H. Endress Foundation.

\appendix

\section{Domain wall states}\label{app:domain}

\subsection{General discussion}\label{app:gen}
Here we give a heuristic discussion of the domain wall ground states referred to in Secs.~\ref{sec:corr} and \ref{sec:entr}. As above, let ${j=1,\ldots,d}$ denote labels for possible positions of the domain wall along the chain, ordered from left to right, and $\ket{\psi_j}$ the corresponding states. We assume that the ground state is well captured within this reduced Hilbert space of domain wall states. In certain limiting cases outlined below, this description is indeed exact. Furthermore, the numerical results of sign-flipping correlation functions, algebraic SF correlations and entanglement entropies indicate that this picture is qualitatively correct even in intermediate parameter regimes. However, one may have to assume that in these regimes the domain wall has a finite width and, therefore, the basis states are not orthogonal [see the discussion below Eq.~\eqref{eq:Slim} in Sec.~\ref{sec:entr}].

An effective Hamiltonian $\hat{H}_\text{eff}$ acting on the domain wall states consist, to zeroth order, of the projection of the full Hamiltonian $\hat{H}$ onto those states. Higher order contributions involve multiple hoppings via virtual states, and, in general, there will be arbitrarily far off-diagonal terms in the effective Hamiltonian. These couplings decay more slowly when the domain wall has a broader width. However, the technicalities of non-orthogonal basis sets do not affect our argument and are, therefore, not discussed (see e.g. Ref.~\cite{Soriano2014}). Rather, we start from the following form of $\hat{H}_\text{eff}$ assuming that the matrix elements ${(\hat{H}_\text{eff})_{ij}}$ depend only on the distance of the two domain walls involved, i.e. 
\begin{equation}
\hat{H}_\text{eff} = \begin{pmatrix}
D       & J_1    & J_2    & \cdots & J_{d-1} \\
J_1     & D      & \ddots & \ddots & \vdots \\
J_2     & \ddots & \ddots &        & J_2 \\
\vdots  & \ddots &        & D      & J_1 \\
J_{d-1} & \cdots & J_2    & J_1    & D \\
\end{pmatrix}.
\label{eq:Heff_general}
\end{equation}
The explicit form of the matrix elements can be calculated by perturbation theory. Deviations are only expected for $i,j$ near the edges in comparison to the width of the domain wall. This is equivalent to a single-particle on a chain where the hopping amplitude between two sites separated by $r$ sites is given by $J_r$ and the onsite energy is given by $D$. The eigenstates $\ket{k}$ of any Hamiltonian of the form of Eq.~\eqref{eq:Heff_general} are given by~\cite{Toeplitz1911} 
\begin{equation}
\ket{k} \propto \sum_j \sin\left(\frac{\pi j k}{d+1}\right) \ket{\psi_j},  k=1,\ldots,d, \label{eq:sine_general}
\end{equation}
where the corresponding energies are
\begin{equation}
E_k = D + 2 \sum_{r} J_r \cos\left( \frac{\pi r k}{d+1} \right).
\end{equation}
For a short-ranged Hamiltonian with $J_1 < 0$, $k=1$ corresponds to the ground state, which motivates our assumption of the sine-like wave function of the domain wall of Eq.~\eqref{eq:dw_wavefunc}. The agreement of this ansatz with our numerical results on the spatial behavior of the entanglement entropies in Fig.~\ref{fig:entr_FSS} suggests that the ground state corresponding to the $(k=1)$-mode is indeed a good description. 

On the other hand, excited states of the effective Hamiltonian are given by modes with higher $k$. Within the reduced Hilbert space, this picture needs to be modified for wave lengths on the order of the width of the domain wall, because the wave function has significant weights on states where Eq.~\eqref{eq:Heff_general} does not hold. Moreover, the reduced Hilbert space itself might not be sufficient, if excitation energies in the effective model are on the order of the bulk excitation gap.

The energy spectrum is gapless, due to long wave-length modes of the domain wall. Even though this can lead to algebraic behavior of the off-diagonals of the single particle density matrix, gapplessness does not imply a bulk superfluid phase. The reason is that the low energy excitations are not bulk properties, but properties of a single defect, i.e. the domain wall. In other words, the off-diagonal long-range order can be understood in terms of $O(L)$ domain wall states, without requiring exponentially many occupation number basis states.

An example for basis states with domain walls of width zero is given in Eq.~\eqref{eq:pattern}. It describes the ground state of the DW phase with boundary condition (A-even) in the limit $V\gg U\gg J$ with effective Hilbert space dimension $d=L/2+1$. The assumption of a zero width domain wall is motivated for intermediate parameter regimes (we used $V=5.0$) by the absence of significant finite size effects in the entanglement entropies [see Fig.~\ref{fig:entr_FSS}(c), even for $L=8$ (not shown)]. In Appendix~\ref{app:eff}, we work out the effective Hamiltonian up to second order hopping processes (via one virtual state) and compare the predictions with exact diagonalization for $L=8$. Because, furthermore, the excited domain wall states are far below the bulk excitation gap of order $U$, we find a good agreement of the low energy spectra.

Similarly, for the low $V$ regime with boundary condition (C) in the limit $U\gg V,J$, appropriate states $\ket{\psi_j}$ are given by Eq.~\eqref{eq:121} with $d=L+1$. These are not strictly domain wall states, but fulfill the same role. However, the finite size scaling of entanglement entropies [see Fig.~\ref{fig:entr_FSS}(a)] indicates that here the assumption of width zero is not applicable in intermediate parameter regimes. Eq. \eqref{eq:HI_dw}, representing domain wall states in the HI phase, is an example of states that do not make this assumption.

\subsection{Example for an effective domain wall Hamiltonian}\label{app:eff}
Here we work out in more detail the effective Hamiltonian on the reduced space of domain wall states for the exemplary case of the DW phase with boundary condition (A-even). We compare the results with exact diagonalization for $L=8$ sites. As mentioned in the main text, in the limit ${V\gg U\gg J}$ the lowest energy states in the occupation number basis are
\begin{equation}
	\ket{\psi_j} = \underbrace{\ket{20\ldots 20}}_{2(j-1)\text{ sites}}\otimes
	\underbrace{\ket{02\ldots 02}}_{L-2(j-1)} \qquad j = 1,\ldots,\frac{L}{2}+1.
	\label{eq:app_effstates}
\end{equation}
with energies $\sim LU/2$.

In order to obtain an effective Hamiltonian of the form of Eq.~\eqref{eq:Heff_general} with nonzero tunneling between our basis states, we consider additional virtual states
\begin{equation}
	\ket{\varphi_l} = \underbrace{\ket{20\ldots 20}}_{2(l-1)\text{ sites}}\otimes\ket{11}\otimes\underbrace{\ket{02\ldots 02}}_{L-2l} \qquad l = 1,\ldots,\frac{L}{2},
	\label{eq:app_virtualstates}
\end{equation}
with energies separated from those of $\ket{\psi_j}$ by $V-U$. We only take into account the lowest order process between $\ket{\psi_j}$ and $\ket{\psi_{j+1}}$, which involves two hoppings, i.e.,
\begin{equation}
	\ket{\psi_j}\rightarrow\ket{\varphi_j}\rightarrow\ket{\psi_{j+1}}.
\end{equation}

The matrix elements of the Hamiltonian given by the states in Eqs.~\eqref{eq:app_effstates} and \eqref{eq:app_virtualstates} are
\begin{equation}
	\begin{split}
	H^a_{ij} &\equiv \bra{\psi_i}\hat{H}\ket{\psi_j} = \frac{L}{2} U \delta_{ij}, \\
	H^b_{ij} &\equiv \bra{\varphi_i}\hat{H}\ket{\varphi_j} = \left[\left(\frac{L}{2}-1\right)U+V \right] \delta_{ij},\\
	W_{ij} &\equiv \bra{\varphi_i}\hat{H}\ket{\psi_{j}} = -\sqrt{2} J (\delta_{ij} + \delta_{i+1, j}).
	\end{split}
\end{equation}
Thus, the Hamiltonian in this subspace can be written as	
\begin{equation}
	\tilde{H} = 
	\begin{pmatrix}
	H^a & W^T \\
	W &  H^b
	\end{pmatrix}
	\in \mathbb{R}^{(L+1)\times (L+1)}.
\end{equation}
When $J \ll U \ll V$, the domain wall states $\ket{\psi_j}$ and the virtual states $\ket{\varphi_j}$ are hybridized by $W$, while $\ket{\psi_j}$ carry almost all the weight in the low energy states. This permits a further restriction of the Hilbert space by ignoring $\ket{\varphi_l}$ (here we follow Chap.~3 of Ref.~\cite{assa}). Using states ${\ket{a}\in\text{span}\{\ket{\psi_j}\}}$ and ${\ket{b}\in\text{span}\{\ket{\varphi_l}\}}$, the Schrödinger equation reads
\begin{equation}
	\begin{split}
	H^a \ket{a} + W^T \ket{b} &= E \ket{a},  \\
	W \ket{a} + H^b \ket{b} &= E \ket{b} . \label{eq:app_Schroed}
	\end{split}
\end{equation}
Focusing on the lower energies ${E\approx LU/2 }$, we can approximate 
\begin{equation}
	E\mathbb{1}_{\frac{L}{2}}- H^b\approx (U-V)\mathbb{1}_{\frac{L}{2}}.
\end{equation}
Then, by solving the second line of Eq.~\eqref{eq:app_Schroed} for $\ket{b}$ and inserting it into the first line, the effective Hamiltonian acting only on $\ket{a}$ attains the tridiagonal form
\begin{equation}
	\begin{split}
	\hat{H}_\text{eff} &= H^{a} + \frac{1}{U-V}W^T W \\
	&= \begin{pmatrix}
	D_\text{edge} & \tilde{J} & 0 & \cdots & 0 \\
	\tilde{J} & D & \tilde{J} & \cdots & 0 \\
	0 & \tilde{J} & \ddots & & \vdots \\
	\vdots & \vdots & & D & \tilde{J} \\
	0 & 0 & \cdots & \tilde{J} & D_\text{edge} \\
	\end{pmatrix} 
	\label{eq:app_Heffeff}
	\end{split}
\end{equation}
with an effective tunneling parameter
\begin{equation}
	\tilde{J} = 2\frac{J^2}{U-V}  < 0 .
	\label{eq:app_TunEff}
\end{equation}
and renormalized diagonal elements
\begin{equation}
	\begin{split}
	D_\text{edge} &= \frac{L}{2} U + 2\frac{J^2}{V-U},  \\
	D &= \frac{L}{2} U + 4\frac{J^2}{V-U}.
	\end{split}
	\label{eq:app_renorm}
\end{equation}
Neglecting the $L$ independent terms in Eq.~\eqref{eq:app_renorm}, which contain the influence of the edge on the domain wall states, the effective Hamiltonian yields the form of Eq.~\eqref{eq:Heff_general}. The eigenvalues for this matrix form are known to be~\cite{Toeplitz1911}
\begin{gather}
	E_k = D + 2\tilde{J}\cos\left(\frac{\pi k}{\frac{L}{2}+2}\right),
	k = 1,\ldots,\frac{L}{2}+1.
	\label{eq:app_Toeplitz}
\end{gather} 
The eigenstates are given by $\ket{k}$ as in Eq.~\eqref{eq:sine_general}.

\begin{figure}[t]
\centering
\includegraphics[width=\columnwidth]{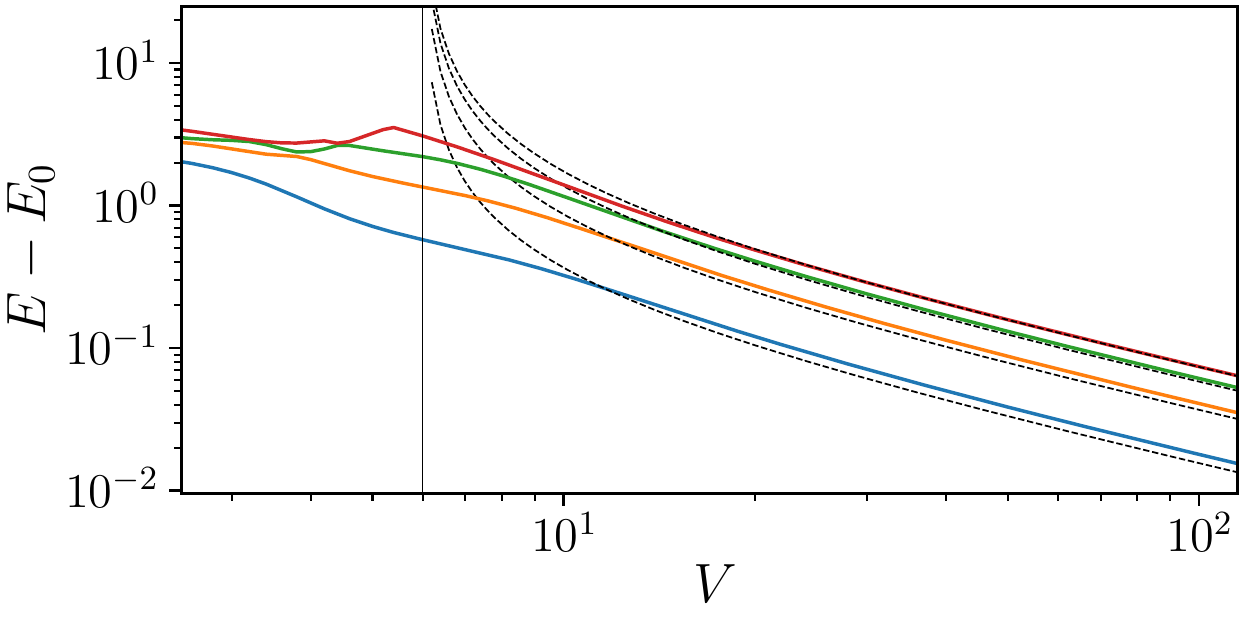}
\caption{Comparison of the exact low energy band (solid lines) with the prediction of the effective theory (dashed lines) for ${U=6}$ and ${L=8}$. Energies are measured relative to the ground state Energy $E_0$.}
\label{fig:app_espec}
\end{figure}

A comparison of the exact low energy band with the prediction of the effective theory confirms the qualitative dependence  of the spectrum on $V$ for ${L=8}$, see Fig.~\ref{fig:app_espec}. This leads us to the conjecture that the spectrum of the DW phase with boundary conditions (A) is actually gapless; the elementary excitations are long wavelength modes of the domain wall. 

The effective theory does not give a good approximation of the energies for ${V\lesssim 10\approx 2U}$ (see Fig.~\ref{fig:app_espec}) and breaks down at $V=U$ due to diverging couplings $\tilde{J}$. However, our DMRG simulations in the previous sections for the DW phase with ${V<U}$ indicate that even in this region the ground state shows qualitatively the same correlations and occupation pattern as our effective theory. An extension of the effective theory by number fluctuations in the domain wall basis states and higher order hopping processes can in principle extend the region of applicability. 

\section{\label{sec:appB}Entanglement enropy of the domain wall state}
In Sec.~\ref{sec:entr} we have compared entanglement entropies that were calculated by DMRG with the sum of a bulk contribution and a generic domain wall contribution. Here we derive the formulas of Eqs.~\eqref{eq:Slim} and \eqref{eq:pexplicit} in the thermodynamic limit. As in Secs.~\ref{sec:entr} and Appendix~\ref{app:domain}), let $\ket{\psi_j}$ (${j=1,\ldots,d=O(L)}$) denote generic domain wall states of finite width, i.e. ${\braket{\psi_i|\psi_j}\to 0}$ for ${|i-j|\to \infty}$. We want to calculate the domain wall entropy for the bipartition at site $l$. For simplicity, let us first assume that the basis states can be written in a product from
\begin{equation}
	\ket{\psi_j} = 
	\begin{cases}
	\ket{L_j}\otimes\ket{R_1} &\text{ for } j< m_l \\
	\ket{L_{m_l}}\otimes\ket{R_{j-m_l+1}} &\text{ for } j \geq m_l.
	\end{cases}, \label{eq:app_leftright}
\end{equation}
where $\ket{L_j}$ ($\ket{R_j}$) represent states on the left (right) block of the chain. The domain wall lies to the left of site $l$ for $j < m_l$, and otherwise to the right. For instance, with the states from Appendix~\ref{app:eff}, this means
\begin{equation}
	\begin{split}
	\ket{L_j}\otimes\ket{R_1} &= \ket{20\ldots 20}\ket{02\ldots 02}\otimes\ket{02\ldots 02} \\
	\ket{L_{m_l}}\otimes\ket{R_{j-m_l+1}} &= \ket{20\ldots 20}\otimes\ket{20\ldots 20}\ket{02\ldots 02},
	\end{split}
\end{equation}
where the domain wall is understood to lie between the sites $2j-2$ and $2j-1$.

If the coefficient of $\ket{\psi_j}$ in the ground state $\ket{G}$ is $\propto w_j$ (e.g. corresponding to the sine-like wave function derived in Appendix~\ref{app:gen}), the weight of the domain wall states in the left block in the ground state is
\begin{equation}
	p_l = \sum_{j < m_l} |\braket{\psi_j|G}|^2 = N^{-1}\sum_{j < m_l}w_j^2\,, \quad N=\sum_{j} w_j^2.
\end{equation}
With this, we can define normalized states on the left and right subsystem
\begin{equation}
	\begin{split}
	\ket{\tilde{L}} &= \frac{1}{\sqrt{p_l N}} \sum_{j < m_l} w_j \ket{L_j} \\ 
	\ket{\tilde{R}} &= \frac{1}{\sqrt{(1-p_l)N}}	\sum_{j \geq m_l} w_j \ket{R_{j-m_l+1}},
	\end{split}
\end{equation}
adding up the states in each of the two cases of Eq.~\eqref{eq:app_leftright}, and express the ground state as
\begin{equation}
	\begin{split}
	\ket{G} &= N^{-\frac{1}{2}} \sum_{j=1}^d w_j \ket{\psi_{j}} \\
	&= N^{-\frac{1}{2}} \left( \sum_{j < m_l} w_j
	\ket{L_j}\ket{R_1} + \sum_{j \geq m_l} w_j \ket{L_{m_l}}\ket{R_{j-m_l+1}} \right) \\
	&= \sqrt{p_l} \ket{\tilde{L}} \ket{R_1} + \sqrt{1-p_l} \ket{L_{m_l}} \ket{\tilde{R}}.
	\end{split}
\end{equation}

Assuming that the basis states are orthogonal, i.e. $\braket{\psi_i|\psi_j}=\delta_{ij}$, the reduced density matrix of the left block simplifies to
\begin{equation}
	\begin{split}
	\rho_{l} &= \text{tr}_{>l}\left\{\ket{G}\bra{G}\right\} \\
	&= \sum_{j \geq m_l} \braket{R_{j-m_l+1}|G}\braket{G|R_{j-m_l+1}} \\
	&= p_l \ket{\tilde{L}}\bra{\tilde{L}} + (1-p_l)
	\ket{L_{m_l}}\bra{L_{m_l}} \\
	&\quad+ \left(\frac{w_{m_l}}{N}\right)^2 \Big( \ket{\tilde{L}}\bra{L_{m_l}} +  \ket{L_{m_l}}\bra{\tilde{L}} \Big),
	\end{split}
\end{equation}
where the off-diagonal elements come exclusively from the terms $\ket{\tilde{L}}\ket{R_1}$ and $\ket{L_{m_l}}\ket{R_1}$. In Appendix~\ref{app:es}, in Eq.~\eqref{eq:app_notdiv4} we see a simple example for a case where the second term is absent, and in Eq.~\eqref{eq:app_div4} an example where it is present. 

Taking the thermodynamic limit $L,d\to \infty$ with fixed $l/L$ and $m_l/d$, we have $w_l/N\to 0$ and thus the reduced density matrix becomes diagonal
\begin{equation}
	\rho_{l} \xrightarrow{L\to\infty} p_l \ket{\tilde{L}}\bra{\tilde{L}} + (1-p_l)
	\ket{L_{m_l}}\bra{L_{m_l}}.
\end{equation}
This obviously gives the domain wall entropy shown in Eq.~\eqref{eq:Slim}. If we do not assume orthogonality, there are larger off-diagonal contributions, due to ${\braket{R_1|R_{j-m_l+1}}\neq 0}$, causing finite size effects as in Figs.~\ref{fig:entr_FSS}~(a)~and~(c). However, these are negligible in the thermodynamic limit.

If we do not assume the product form of Eq.~\eqref{eq:app_leftright}, we have to write the states as a Schmidt-decomposition
\begin{equation}
	\ket{\psi_j} = 
	\begin{cases}
	\sum_\alpha \lambda_j^\alpha \ket{L_j^\alpha}\otimes\ket{R_1^\alpha} &\text{ for } j < m_l \\
	\sum_\alpha \lambda_j^\alpha \ket{L_{m_l}^\alpha}\otimes\ket{R_{j-m_l+1}^\alpha} &\text{ for } j \geq m_l.
	\end{cases}, \label{eq:app_leftright_Schmidt}
\end{equation}
where $\{(\lambda_j^\alpha)^2:\alpha\}$ is the entanglement spectrum of the bipartition of $\ket{\psi_j}$ at site $l$. For $|j-m_l|$ sufficiently large, i.e. when the domain wall is sufficiently far from site $l$, this corresponds to the entanglement spectrum of the underlying bulk phase ($\lambda_j^\alpha\to \lambda^\alpha$). Going through the same calculation as above and ignoring contributions where the domain wall is close to $l$ (as the corresponding weights $w_j/N$ vanish in the thermodynamic limit), we first define
\begin{equation}
	\begin{split}
	\ket{\tilde{L}^\alpha} &= \frac{1}{\sqrt{p_l N}} \sum_{j < m_l} w_j \ket{L_j^\alpha} \\ 
	\ket{\tilde{R}^\alpha} &= \frac{1}{\sqrt{(1-p_l)N}}	\sum_{j \geq m_l} w_j \ket{R_{j-m_l+1}^\alpha},
	\end{split}
\end{equation}
and then arrive at
\begin{equation}
	\rho_{l} \xrightarrow{L\to\infty} p_l \sum_\alpha \lambda^\alpha \ket{\tilde{L}^\alpha}\bra{\tilde{L}^\alpha} + (1-p_l) \sum_\alpha \lambda^\alpha
	\ket{L_{m_l}^\alpha}\bra{L_{m_l}^\alpha}.
\end{equation}
From this it follows that the entropy can be written as
\begin{equation}
	\begin{split}
	S &= -\sum_\alpha p_l\lambda^\alpha \ln(p_l\lambda^\alpha) + (1-p_l)\lambda^\alpha \ln((1-p_l)\lambda^\alpha) \\
	&= -\sum_\alpha \lambda^\alpha \ln(\lambda^\alpha) - p_l \ln(p_l) - (1-p_l) \ln(1-p_l).
	\end{split}
\end{equation}
The first term is the entropy of the underlying bulk phase, while the rest is gives domain wall contribution. This explains the results from Sec.~\ref{sec:entr}.

\section{Entanglement spectra}\label{app:es}

\begin{figure}[t!]
	\centering
	\includegraphics[width=\columnwidth]{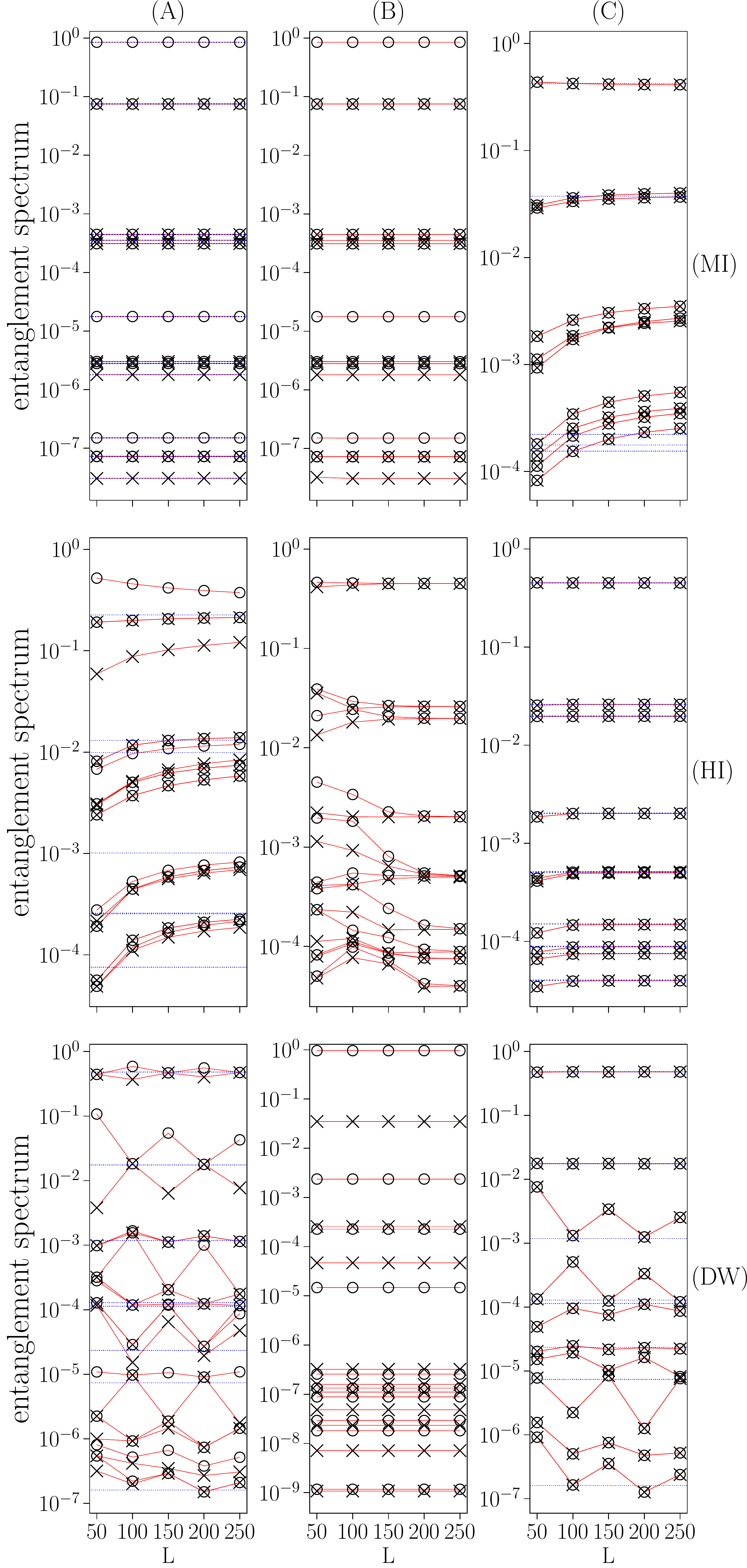}
	\caption{Finite size scaling of the $20$ largest values of the entanglment spectra. Columns correspond to boundary condition (A), (B) and (C). Rows represent points deep in the MI (${V=1.0}$), HI (${V=3.7}$) and DW (${V=5.0}$) phases. Dotted blue lines indicate the bulk values from (B) for MI-(A) and HI-(C) and half those values otherwise. Circles and crosses as labels alternate as an aid to the eye.}
	\label{fig:app_entSpec}
\end{figure}

As a supplement to our analysis of entanglement entropies in Sec.~\ref{sec:entr}, here we show the most significant values of the full entanglement spectra for the symmetric bipartition at $L/2$. First, this offers a closer look at the system size dependence. Second, we can study degeneracies; in particular, the HI phase is known to have a doubly degenerate entanglement spectrum~\cite{Pollmann2010,Deng2011}. This is a footprint of the HI being a so called symmetry-protected topological phase: as long as a certain symmetry of the Hamiltonian (e.g. lattice inversion for the HI phase) is not broken by a change of its parameters, the degeneracies can only be removed through a phase transition.

In the present case, we have to deal with an additional, trivial type of degeneracy due to the fact that lattice inversion symmetry is present in all ground states. Let $\rho_l$ denote the block of the reduced density matrix of the left half of the system with particle number $l$. As a consequence of the inversion symmetry, the blocks fulfill the relation $\rho_l=\rho_{L-l}$~\cite{Deng2011}. For total particle number $N=L$, as in case of the boundary conditions (A), this implies a partial double degeneracy of the entanglement spectrum. Only the block $\rho_{L/2}$ has nondegenerate counterpart. On the other hand, for boundary condition (C) with $N=L+1$ we have a complete double degeneracy, because the block $\rho_{L/2}=\rho_{L/2+1}$ now has a counterpart. With boundary condition (B), where we used edge potentials to break the inversion symmetry, we do not expect such degeneracies.

In Fig.~\ref{fig:app_entSpec} we show entanglement spectra for boundary condition (A-even), (B) and (C), respectively. Each of the figures contains data on one ground state in the MI ($V=1.0$), HI ($V=3.7$) and DW ($V=5.0$). The first thing we note are the partial double degeneracy with boundary condition (A) [first column] and the complete double degeneracy with boundary condition (C) [third column], which are present independently of the phase. These degeneracies are due to the block symmetry described above. In addition, in the DW phase of BC (A) the values of the entanglement spectrum oscillate (see bottom left panel). For the largest two eigenvalues, this can be understood in terms of the effective picture described in Appendix~\ref{app:eff}: If $L$ is divisible by $4$, the basis states can decomposed as
\begin{equation}
	\begin{split}
	&\underbrace{\ket{20\ldots 20}\ket{02\ldots 02}}_{L/2\text{ particles}}\otimes\underbrace{\ket{02\ldots 02}}_{L/2 \text{ particles}}, \\
	&\underbrace{\ket{20\ldots 20}}_{L/2\text{ particles}}\otimes\underbrace{\ket{20\ldots 20}\ket{02\ldots 02}}_{L/2 \text{ particles}},
	\label{eq:app_div4}
	\end{split}
\end{equation}
where the $\otimes$-sign indicates the middle of the chain and the domain wall is either on the left (upper line) or the right (lower line) of the middle. If $L$ is not divisible by $4$, the basis states are decomposed as
\begin{equation}
	\begin{split}
	&\underbrace{\ket{20\ldots 20}\ket{02\ldots 020}}_{L/2-1\text{ particles}}\otimes\underbrace{\ket{202\ldots 02}}_{L/2+1 \text{ particles}} \\
	&\underbrace{\ket{20\ldots 202}}_{L/2+1\text{ particles}}\otimes\underbrace{\ket{020\ldots 20}\ket{02\ldots 02}}_{L/2-1 \text{ particles}}.
	\label{eq:app_notdiv4}
	\end{split}	
\end{equation}
In the case of Eq.~\eqref{eq:app_notdiv4}, these states give two degenerate contributions $\rho_{L/2\pm 1}$ to the reduced density matrix. Eq.~\eqref{eq:app_div4} makes only one contribution $\rho_{L/2}$, which does not imply degeneracies. 

We note that the entanglement spectrum converges in $L$ for boundary condition (B) [cf. middle column]. We consider these to be the bulk values. Here, despite the explicitly broken inversion symmetry, the HI is characterized by a double degeneracy and the MI by a partial double degeneracy. A comparison with boundary conditions (A) and (C) shows that in cases with no domain wall the values coincide [i.e. MI with (A) and HI with (C)]. On the other hand, if there is a domain wall [i.e. MI with (C), HI with (A), DW for both (A) and (C)], many values converge slowly in the system size (we checked this to be consistent with a power law, not shown). In these cases, the thermodynamic limit is consistent with half the bulk value, which implies the behavior of entanglement entropies found in Sec.~\ref{sec:entr} (see blue lines in Fig.~\ref{fig:app_entSpec}). In particular, the largest value of the sign-flipping HI with boundary condition (A) appears to be fourfold instead of twofold degenerate at half the bulk value in the thermodynamic limit. This can be understood as a combination of the topologically protected double degeneracy and the influence of the domain wall. One exception are the values below the sixth-largest in the MI with boundary condition (C), where the extra boson is delocalized. These do not have a counterpart in the bulk spectrum.

\nocite{*}


\begin{thebibliography}{33}%
	\makeatletter
	\providecommand \@ifxundefined [1]{%
		\@ifx{#1\undefined}
	}%
	\providecommand \@ifnum [1]{%
		\ifnum #1\expandafter \@firstoftwo
		\else \expandafter \@secondoftwo
		\fi
	}%
	\providecommand \@ifx [1]{%
		\ifx #1\expandafter \@firstoftwo
		\else \expandafter \@secondoftwo
		\fi
	}%
	\providecommand \natexlab [1]{#1}%
	\providecommand \enquote  [1]{``#1''}%
	\providecommand \bibnamefont  [1]{#1}%
	\providecommand \bibfnamefont [1]{#1}%
	\providecommand \citenamefont [1]{#1}%
	\providecommand \href@noop [0]{\@secondoftwo}%
	\providecommand \href [0]{\begingroup \@sanitize@url \@href}%
	\providecommand \@href[1]{\@@startlink{#1}\@@href}%
	\providecommand \@@href[1]{\endgroup#1\@@endlink}%
	\providecommand \@sanitize@url [0]{\catcode `\\12\catcode `\$12\catcode
		`\&12\catcode `\#12\catcode `\^12\catcode `\_12\catcode `\%12\relax}%
	\providecommand \@@startlink[1]{}%
	\providecommand \@@endlink[0]{}%
	\providecommand \url  [0]{\begingroup\@sanitize@url \@url }%
	\providecommand \@url [1]{\endgroup\@href {#1}{\urlprefix }}%
	\providecommand \urlprefix  [0]{URL }%
	\providecommand \Eprint [0]{\href }%
	\providecommand \doibase [0]{https://doi.org/}%
	\providecommand \selectlanguage [0]{\@gobble}%
	\providecommand \bibinfo  [0]{\@secondoftwo}%
	\providecommand \bibfield  [0]{\@secondoftwo}%
	\providecommand \translation [1]{[#1]}%
	\providecommand \BibitemOpen [0]{}%
	\providecommand \bibitemStop [0]{}%
	\providecommand \bibitemNoStop [0]{.\EOS\space}%
	\providecommand \EOS [0]{\spacefactor3000\relax}%
	\providecommand \BibitemShut  [1]{\csname bibitem#1\endcsname}%
	\let\auto@bib@innerbib\@empty
	\bibitem [{\citenamefont {Bloch}\ and\ \citenamefont
		{Immanuel}(2005)}]{Bloch2008}%
	\BibitemOpen
	\bibfield  {author} {\bibinfo {author} {\bibfnamefont {I.}~\bibnamefont
			{Bloch}}\ and\ \bibinfo {author} {\bibnamefont {Immanuel}},\ }\bibfield
	{title} {\bibinfo {title} {{Ultracold quantum gases in optical lattices}},\
	}\href {https://doi.org/10.1038/nphys138} {\bibfield  {journal} {\bibinfo
			{journal} {Nat. Phys.}\ }\textbf {\bibinfo {volume} {1}},\ \bibinfo {pages}
		{23} (\bibinfo {year} {2005})}\BibitemShut {NoStop}%
	\bibitem [{\citenamefont {Lewenstein}\ \emph {et~al.}(2007)\citenamefont
		{Lewenstein}, \citenamefont {Sanpera}, \citenamefont {Ahufinger},
		\citenamefont {Damski}, \citenamefont {Sen(De)},\ and\ \citenamefont
		{Sen}}]{Lewenstein2007}%
	\BibitemOpen
	\bibfield  {author} {\bibinfo {author} {\bibfnamefont {M.}~\bibnamefont
			{Lewenstein}}, \bibinfo {author} {\bibfnamefont {A.}~\bibnamefont {Sanpera}},
		\bibinfo {author} {\bibfnamefont {V.}~\bibnamefont {Ahufinger}}, \bibinfo
		{author} {\bibfnamefont {B.}~\bibnamefont {Damski}}, \bibinfo {author}
		{\bibfnamefont {A.}~\bibnamefont {Sen(De)}},\ and\ \bibinfo {author}
		{\bibfnamefont {U.}~\bibnamefont {Sen}},\ }\bibfield  {title} {\bibinfo
		{title} {{Ultracold atomic gases in optical lattices: mimicking condensed
				matter physics and beyond}},\ }\href
	{https://doi.org/10.1080/00018730701223200} {\bibfield  {journal} {\bibinfo
			{journal} {Adv. Phys.}\ }\textbf {\bibinfo {volume} {56}},\ \bibinfo {pages}
		{243} (\bibinfo {year} {2007})}\BibitemShut {NoStop}%
	\bibitem [{\citenamefont {Jaksch}\ and\ \citenamefont
		{Zoller}(2005)}]{Jaksch2005}%
	\BibitemOpen
	\bibfield  {author} {\bibinfo {author} {\bibfnamefont {D.}~\bibnamefont
			{Jaksch}}\ and\ \bibinfo {author} {\bibfnamefont {P.}~\bibnamefont
			{Zoller}},\ }\bibfield  {title} {\bibinfo {title} {{The cold atom Hubbard
				toolbox}},\ }\href {https://doi.org/10.1016/j.aop.2004.09.010} {\bibfield
		{journal} {\bibinfo  {journal} {Ann. Phys. (N. Y).}\ }\textbf {\bibinfo
			{volume} {315}},\ \bibinfo {pages} {52} (\bibinfo {year} {2005})}\BibitemShut
	{NoStop}%
	\bibitem [{\citenamefont {Lahaye}\ \emph {et~al.}(2009)\citenamefont {Lahaye},
		\citenamefont {Menotti}, \citenamefont {Santos}, \citenamefont {Lewenstein},\
		and\ \citenamefont {Pfau}}]{Lahaye2009}%
	\BibitemOpen
	\bibfield  {author} {\bibinfo {author} {\bibfnamefont {T.}~\bibnamefont
			{Lahaye}}, \bibinfo {author} {\bibfnamefont {C.}~\bibnamefont {Menotti}},
		\bibinfo {author} {\bibfnamefont {L.}~\bibnamefont {Santos}}, \bibinfo
		{author} {\bibfnamefont {M.}~\bibnamefont {Lewenstein}},\ and\ \bibinfo
		{author} {\bibfnamefont {T.}~\bibnamefont {Pfau}},\ }\bibfield  {title}
	{\bibinfo {title} {{The physics of dipolar bosonic quantum gases}},\ }\href
	{https://doi.org/10.1088/0034-4885/72/12/126401} {\bibfield  {journal}
		{\bibinfo  {journal} {Reports Prog. Phys.}\ }\textbf {\bibinfo {volume}
			{72}},\ \bibinfo {pages} {126401} (\bibinfo {year} {2009})}\BibitemShut
	{NoStop}%
	\bibitem [{\citenamefont {Baier}\ \emph {et~al.}(2016)\citenamefont {Baier},
		\citenamefont {Mark}, \citenamefont {Petter}, \citenamefont {Aikawa},
		\citenamefont {Chomaz}, \citenamefont {Cai}, \citenamefont {Baranov},
		\citenamefont {Zoller},\ and\ \citenamefont {Ferlaino}}]{Baier2016}%
	\BibitemOpen
	\bibfield  {author} {\bibinfo {author} {\bibfnamefont {S.}~\bibnamefont
			{Baier}}, \bibinfo {author} {\bibfnamefont {M.~J.}\ \bibnamefont {Mark}},
		\bibinfo {author} {\bibfnamefont {D.}~\bibnamefont {Petter}}, \bibinfo
		{author} {\bibfnamefont {K.}~\bibnamefont {Aikawa}}, \bibinfo {author}
		{\bibfnamefont {L.}~\bibnamefont {Chomaz}}, \bibinfo {author} {\bibfnamefont
			{Z.}~\bibnamefont {Cai}}, \bibinfo {author} {\bibfnamefont {M.}~\bibnamefont
			{Baranov}}, \bibinfo {author} {\bibfnamefont {P.}~\bibnamefont {Zoller}},\
		and\ \bibinfo {author} {\bibfnamefont {F.}~\bibnamefont {Ferlaino}},\
	}\bibfield  {title} {\bibinfo {title} {{Extended Bose-Hubbard models with
				ultracold magnetic atoms}},\ }\href {https://doi.org/10.1126/science.aac9812}
	{\bibfield  {journal} {\bibinfo  {journal} {Science}\ }\textbf {\bibinfo
			{volume} {352}},\ \bibinfo {pages} {201} (\bibinfo {year}
		{2016})}\BibitemShut {NoStop}%
	\bibitem [{\citenamefont {K{\"{u}}hner}\ \emph {et~al.}(2000)\citenamefont
		{K{\"{u}}hner}, \citenamefont {White},\ and\ \citenamefont
		{Monien}}]{Kuehner2000}%
	\BibitemOpen
	\bibfield  {author} {\bibinfo {author} {\bibfnamefont {T.~D.}\ \bibnamefont
			{K{\"{u}}hner}}, \bibinfo {author} {\bibfnamefont {S.~R.}\ \bibnamefont
			{White}},\ and\ \bibinfo {author} {\bibfnamefont {H.}~\bibnamefont
			{Monien}},\ }\bibfield  {title} {\bibinfo {title} {{One-dimensional
				Bose-Hubbard model with nearest-neighbor interaction}},\ }\href
	{https://doi.org/10.1103/PhysRevB.61.12474} {\bibfield  {journal} {\bibinfo
			{journal} {Phys. Rev. B}\ }\textbf {\bibinfo {volume} {61}},\ \bibinfo
		{pages} {12474} (\bibinfo {year} {2000})}\BibitemShut {NoStop}%
	\bibitem [{\citenamefont {{Dalla Torre}}\ \emph {et~al.}(2006)\citenamefont
		{{Dalla Torre}}, \citenamefont {Berg},\ and\ \citenamefont
		{Altman}}]{DallaTorre2006}%
	\BibitemOpen
	\bibfield  {author} {\bibinfo {author} {\bibfnamefont {E.~G.}\ \bibnamefont
			{{Dalla Torre}}}, \bibinfo {author} {\bibfnamefont {E.}~\bibnamefont
			{Berg}},\ and\ \bibinfo {author} {\bibfnamefont {E.}~\bibnamefont {Altman}},\
	}\bibfield  {title} {\bibinfo {title} {{Hidden Order in 1D Bose
				Insulators}},\ }\href {https://doi.org/10.1103/PhysRevLett.97.260401}
	{\bibfield  {journal} {\bibinfo  {journal} {Phys. Rev. Lett.}\ }\textbf
		{\bibinfo {volume} {97}},\ \bibinfo {pages} {260401} (\bibinfo {year}
		{2006})}\BibitemShut {NoStop}%
	\bibitem [{\citenamefont {Berg}\ \emph {et~al.}(2008)\citenamefont {Berg},
		\citenamefont {{Dalla Torre}}, \citenamefont {Giamarchi},\ and\ \citenamefont
		{Altman}}]{Berg2008}%
	\BibitemOpen
	\bibfield  {author} {\bibinfo {author} {\bibfnamefont {E.}~\bibnamefont
			{Berg}}, \bibinfo {author} {\bibfnamefont {E.~G.}\ \bibnamefont {{Dalla
					Torre}}}, \bibinfo {author} {\bibfnamefont {T.}~\bibnamefont {Giamarchi}},\
		and\ \bibinfo {author} {\bibfnamefont {E.}~\bibnamefont {Altman}},\
	}\bibfield  {title} {\bibinfo {title} {{Rise and fall of hidden string order
				of lattice bosons}},\ }\href {https://doi.org/10.1103/PhysRevB.77.245119}
	{\bibfield  {journal} {\bibinfo  {journal} {Phys. Rev. B}\ }\textbf {\bibinfo
			{volume} {77}},\ \bibinfo {pages} {245119} (\bibinfo {year}
		{2008})}\BibitemShut {NoStop}%
	\bibitem [{\citenamefont {Ejima}\ \emph {et~al.}(2014)\citenamefont {Ejima},
		\citenamefont {Lange},\ and\ \citenamefont {Fehske}}]{Ejima2014PRL}%
	\BibitemOpen
	\bibfield  {author} {\bibinfo {author} {\bibfnamefont {S.}~\bibnamefont
			{Ejima}}, \bibinfo {author} {\bibfnamefont {F.}~\bibnamefont {Lange}},\ and\
		\bibinfo {author} {\bibfnamefont {H.}~\bibnamefont {Fehske}},\ }\bibfield
	{title} {\bibinfo {title} {{Spectral and Entanglement Properties of the
				Bosonic Haldane Insulator}},\ }\href
	{https://doi.org/10.1103/PhysRevLett.113.020401} {\bibfield  {journal}
		{\bibinfo  {journal} {Phys. Rev. Lett.}\ }\textbf {\bibinfo {volume} {113}},\
		\bibinfo {pages} {020401} (\bibinfo {year} {2014})}\BibitemShut {NoStop}%
	\bibitem [{\citenamefont {Pollmann}\ \emph {et~al.}(2010)\citenamefont
		{Pollmann}, \citenamefont {Turner}, \citenamefont {Berg},\ and\ \citenamefont
		{Oshikawa}}]{Pollmann2010}%
	\BibitemOpen
	\bibfield  {author} {\bibinfo {author} {\bibfnamefont {F.}~\bibnamefont
			{Pollmann}}, \bibinfo {author} {\bibfnamefont {A.~M.}\ \bibnamefont
			{Turner}}, \bibinfo {author} {\bibfnamefont {E.}~\bibnamefont {Berg}},\ and\
		\bibinfo {author} {\bibfnamefont {M.}~\bibnamefont {Oshikawa}},\ }\bibfield
	{title} {\bibinfo {title} {{Entanglement spectrum of a topological phase in
				one dimension}},\ }\href {https://doi.org/10.1103/PhysRevB.81.064439}
	{\bibfield  {journal} {\bibinfo  {journal} {Phys. Rev. B}\ }\textbf {\bibinfo
			{volume} {81}},\ \bibinfo {pages} {064439} (\bibinfo {year}
		{2010})}\BibitemShut {NoStop}%
	\bibitem [{\citenamefont {Muldoon}\ \emph {et~al.}(2012)\citenamefont
		{Muldoon}, \citenamefont {Brandt}, \citenamefont {Dong}, \citenamefont
		{Stuart}, \citenamefont {Brainis}, \citenamefont {Himsworth},\ and\
		\citenamefont {Kuhn}}]{Muldoon2012}%
	\BibitemOpen
	\bibfield  {author} {\bibinfo {author} {\bibfnamefont {C.}~\bibnamefont
			{Muldoon}}, \bibinfo {author} {\bibfnamefont {L.}~\bibnamefont {Brandt}},
		\bibinfo {author} {\bibfnamefont {J.}~\bibnamefont {Dong}}, \bibinfo {author}
		{\bibfnamefont {D.}~\bibnamefont {Stuart}}, \bibinfo {author} {\bibfnamefont
			{E.}~\bibnamefont {Brainis}}, \bibinfo {author} {\bibfnamefont
			{M.}~\bibnamefont {Himsworth}},\ and\ \bibinfo {author} {\bibfnamefont
			{A.}~\bibnamefont {Kuhn}},\ }\bibfield  {title} {\bibinfo {title} {{Control
				and manipulation of cold atoms in optical tweezers}},\ }\href
	{https://doi.org/10.1088/1367-2630/14/7/073051} {\bibfield  {journal}
		{\bibinfo  {journal} {New J. Phys.}\ }\textbf {\bibinfo {volume} {14}},\
		\bibinfo {pages} {073051} (\bibinfo {year} {2012})}\BibitemShut {NoStop}%
	\bibitem [{\citenamefont {Jaschke}\ \emph {et~al.}(2017)\citenamefont
		{Jaschke}, \citenamefont {{L. Wall}},\ and\ \citenamefont {{D.
				Carr}}}]{openMPS}%
	\BibitemOpen
	\bibfield  {author} {\bibinfo {author} {\bibfnamefont {D.}~\bibnamefont
			{Jaschke}}, \bibinfo {author} {\bibfnamefont {M.}~\bibnamefont {{L. Wall}}},\
		and\ \bibinfo {author} {\bibfnamefont {L.}~\bibnamefont {{D. Carr}}},\
	}\bibfield  {title} {\bibinfo {title} {{Open source Matrix Product States:
				Opening ways to simulate entangled many-body quantum systems in one
				dimension}},\ }\href {https://doi.org/10.1016/j.cpc.2017.12.015} {\bibfield
		{journal} {\bibinfo  {journal} {Comput. Phys. Commun.}\ }\textbf {\bibinfo
			{volume} {225}},\ \bibinfo {pages} {59} (\bibinfo {year} {2017})}\BibitemShut
	{NoStop}%
	\bibitem [{\citenamefont {Chaikin}\ and\ \citenamefont
		{Lubensky}(1995)}]{Chaikin1995}%
	\BibitemOpen
	\bibfield  {author} {\bibinfo {author} {\bibfnamefont {P.~M.}\ \bibnamefont
			{Chaikin}}\ and\ \bibinfo {author} {\bibfnamefont {T.~C.}\ \bibnamefont
			{Lubensky}},\ }\href {https://doi.org/10.1017/CBO9780511813467} {\emph
		{\bibinfo {title} {{Principles of Condensed Matter Physics}}}}\ (\bibinfo
	{publisher} {Cambridge University Press},\ \bibinfo {address} {Cambridge,
		England},\ \bibinfo {year} {1995})\BibitemShut {NoStop}%
	\bibitem [{\citenamefont {{Dalla Torre}}(2013)}]{DallaTorre2013}%
	\BibitemOpen
	\bibfield  {author} {\bibinfo {author} {\bibfnamefont {E.~G.}\ \bibnamefont
			{{Dalla Torre}}},\ }\bibfield  {title} {\bibinfo {title} {{Dynamical probing
				of a topological phase of bosons in one dimension}},\ }\href
	{https://doi.org/10.1088/0953-4075/46/8/085303} {\bibfield  {journal}
		{\bibinfo  {journal} {J. Phys. B At. Mol. Opt. Phys.}\ }\textbf {\bibinfo
			{volume} {46}},\ \bibinfo {pages} {085303} (\bibinfo {year}
		{2013})}\BibitemShut {NoStop}%
	\bibitem [{\citenamefont {Rossini}\ and\ \citenamefont
		{Fazio}(2012)}]{Rossini2012}%
	\BibitemOpen
	\bibfield  {author} {\bibinfo {author} {\bibfnamefont {D.}~\bibnamefont
			{Rossini}}\ and\ \bibinfo {author} {\bibfnamefont {R.}~\bibnamefont
			{Fazio}},\ }\bibfield  {title} {\bibinfo {title} {{Phase diagram of the
				extended Bose–Hubbard model}},\ }\href
	{https://doi.org/10.1088/1367-2630/14/6/065012} {\bibfield  {journal}
		{\bibinfo  {journal} {New J. Phys.}\ }\textbf {\bibinfo {volume} {14}},\
		\bibinfo {pages} {065012} (\bibinfo {year} {2012})}\BibitemShut {NoStop}%
	\bibitem [{\citenamefont {Batrouni}\ \emph {et~al.}(2013)\citenamefont
		{Batrouni}, \citenamefont {Scalettar}, \citenamefont {Rousseau},\ and\
		\citenamefont {Gr{\'{e}}maud}}]{Batrouni2013}%
	\BibitemOpen
	\bibfield  {author} {\bibinfo {author} {\bibfnamefont {G.~G.}\ \bibnamefont
			{Batrouni}}, \bibinfo {author} {\bibfnamefont {R.~T.}\ \bibnamefont
			{Scalettar}}, \bibinfo {author} {\bibfnamefont {V.~G.}\ \bibnamefont
			{Rousseau}},\ and\ \bibinfo {author} {\bibfnamefont {B.}~\bibnamefont
			{Gr{\'{e}}maud}},\ }\bibfield  {title} {\bibinfo {title} {{Competing
				Supersolid and Haldane Insulator Phases in the Extended One-Dimensional
				Bosonic Hubbard Model}},\ }\href
	{https://doi.org/10.1103/PhysRevLett.110.265303} {\bibfield  {journal}
		{\bibinfo  {journal} {Phys. Rev. Lett.}\ }\textbf {\bibinfo {volume} {110}},\
		\bibinfo {pages} {265303} (\bibinfo {year} {2013})}\BibitemShut {NoStop}%
	\bibitem [{\citenamefont {Batrouni}\ \emph {et~al.}(2014)\citenamefont
		{Batrouni}, \citenamefont {Rousseau}, \citenamefont {Scalettar},\ and\
		\citenamefont {Gr{\'{e}}maud}}]{Batrouni2014}%
	\BibitemOpen
	\bibfield  {author} {\bibinfo {author} {\bibfnamefont {G.~G.}\ \bibnamefont
			{Batrouni}}, \bibinfo {author} {\bibfnamefont {V.~G.}\ \bibnamefont
			{Rousseau}}, \bibinfo {author} {\bibfnamefont {R.~T.}\ \bibnamefont
			{Scalettar}},\ and\ \bibinfo {author} {\bibfnamefont {B.}~\bibnamefont
			{Gr{\'{e}}maud}},\ }\bibfield  {title} {\bibinfo {title} {{Competing phases,
				phase separation, and coexistence in the extended one-dimensional bosonic
				Hubbard model}},\ }\href {https://doi.org/10.1103/PhysRevB.90.205123}
	{\bibfield  {journal} {\bibinfo  {journal} {Phys. Rev. B}\ }\textbf {\bibinfo
			{volume} {90}},\ \bibinfo {pages} {205123} (\bibinfo {year}
		{2014})}\BibitemShut {NoStop}%
	\bibitem [{\citenamefont {Batrouni}\ \emph {et~al.}(2015)\citenamefont
		{Batrouni}, \citenamefont {Rousseau}, \citenamefont {Scalettar},\ and\
		\citenamefont {Gr{\'{e}}maud}}]{Batrouni2015}%
	\BibitemOpen
	\bibfield  {author} {\bibinfo {author} {\bibfnamefont {G.~G.}\ \bibnamefont
			{Batrouni}}, \bibinfo {author} {\bibfnamefont {V.~G.}\ \bibnamefont
			{Rousseau}}, \bibinfo {author} {\bibfnamefont {R.~T.}\ \bibnamefont
			{Scalettar}},\ and\ \bibinfo {author} {\bibfnamefont {B.}~\bibnamefont
			{Gr{\'{e}}maud}},\ }\bibfield  {title} {\bibinfo {title} {{Competition
				between the Haldane insulator, superfluid and supersolid phases in the
				one-dimensional Bosonic Hubbard Model}},\ }\href
	{https://doi.org/10.1088/1742-6596/640/1/012042} {\bibfield  {journal}
		{\bibinfo  {journal} {J. Phys. Conf. Ser.}\ }\textbf {\bibinfo {volume}
			{640}},\ \bibinfo {pages} {012042} (\bibinfo {year} {2015})}\BibitemShut
	{NoStop}%
	\bibitem [{\citenamefont {Kollath}\ \emph {et~al.}(2010)\citenamefont
		{Kollath}, \citenamefont {Roux}, \citenamefont {Biroli},\ and\ \citenamefont
		{L{\"{a}}uchli}}]{Kollath2010}%
	\BibitemOpen
	\bibfield  {author} {\bibinfo {author} {\bibfnamefont {C.}~\bibnamefont
			{Kollath}}, \bibinfo {author} {\bibfnamefont {G.}~\bibnamefont {Roux}},
		\bibinfo {author} {\bibfnamefont {G.}~\bibnamefont {Biroli}},\ and\ \bibinfo
		{author} {\bibfnamefont {A.~M.}\ \bibnamefont {L{\"{a}}uchli}},\ }\bibfield
	{title} {\bibinfo {title} {{Statistical properties of the spectrum of the
				extended Bose–Hubbard model}},\ }\href
	{https://doi.org/10.1088/1742-5468/2010/08/P08011} {\bibinfo  {journal} {J.
			Stat. Mech.: Theory Exp.}\ ,\ \bibinfo {pages} {P08011}}\BibitemShut
	{NoStop}%
	\bibitem [{\citenamefont {Gr{\'{e}}maud}\ and\ \citenamefont
		{Batrouni}(2016)}]{Gremaud2016}%
	\BibitemOpen
	\bibfield  {journal} {  }\bibfield  {author} {\bibinfo {author} {\bibfnamefont
			{B.}~\bibnamefont {Gr{\'{e}}maud}}\ and\ \bibinfo {author} {\bibfnamefont
			{G.~G.}\ \bibnamefont {Batrouni}},\ }\bibfield  {title} {\bibinfo {title}
		{{Excitation and dynamics in the extended Bose-Hubbard model}},\ }\href
	{https://doi.org/10.1103/PhysRevB.93.035108} {\bibfield  {journal} {\bibinfo
			{journal} {Phys. Rev. B}\ }\textbf {\bibinfo {volume} {93}},\ \bibinfo
		{pages} {035108} (\bibinfo {year} {2016})}\BibitemShut {NoStop}%
	\bibitem [{\citenamefont {Deng}\ and\ \citenamefont {Santos}(2011)}]{Deng2011}%
	\BibitemOpen
	\bibfield  {author} {\bibinfo {author} {\bibfnamefont {X.}~\bibnamefont
			{Deng}}\ and\ \bibinfo {author} {\bibfnamefont {L.}~\bibnamefont {Santos}},\
	}\bibfield  {title} {\bibinfo {title} {{Entanglement spectrum of
				one-dimensional extended Bose-Hubbard models}},\ }\href
	{https://doi.org/10.1103/PhysRevB.84.085138} {\bibfield  {journal} {\bibinfo
			{journal} {Phys. Rev. B}\ }\textbf {\bibinfo {volume} {84}},\ \bibinfo
		{pages} {085138} (\bibinfo {year} {2011})}\BibitemShut {NoStop}%
	\bibitem [{\citenamefont {zu~M{\"{u}}nster}\ \emph {et~al.}(2014)\citenamefont
		{zu~M{\"{u}}nster}, \citenamefont {Gebhard}, \citenamefont {Ejima},\ and\
		\citenamefont {Fehske}}]{Ejima2014PRA}%
	\BibitemOpen
	\bibfield  {author} {\bibinfo {author} {\bibfnamefont {K.}~\bibnamefont
			{zu~M{\"{u}}nster}}, \bibinfo {author} {\bibfnamefont {F.}~\bibnamefont
			{Gebhard}}, \bibinfo {author} {\bibfnamefont {S.}~\bibnamefont {Ejima}},\
		and\ \bibinfo {author} {\bibfnamefont {H.}~\bibnamefont {Fehske}},\
	}\bibfield  {title} {\bibinfo {title} {{Dynamical correlation functions for
				the one-dimensional Bose-Hubbard insulator}},\ }\href
	{https://doi.org/10.1103/PhysRevA.89.063623} {\bibfield  {journal} {\bibinfo
			{journal} {Phys. Rev. A}\ }\textbf {\bibinfo {volume} {89}},\ \bibinfo
		{pages} {063623} (\bibinfo {year} {2014})}\BibitemShut {NoStop}%
	\bibitem [{\citenamefont {Kurdestany}\ \emph {et~al.}(2014)\citenamefont
		{Kurdestany}, \citenamefont {Pai}, \citenamefont {Mukerjee},\ and\
		\citenamefont {Pandit}}]{Kurdestany2014}%
	\BibitemOpen
	\bibfield  {author} {\bibinfo {author} {\bibfnamefont {J.~M.}\ \bibnamefont
			{Kurdestany}}, \bibinfo {author} {\bibfnamefont {R.~V.}\ \bibnamefont {Pai}},
		\bibinfo {author} {\bibfnamefont {S.}~\bibnamefont {Mukerjee}},\ and\
		\bibinfo {author} {\bibfnamefont {R.}~\bibnamefont {Pandit}},\ }\bibfield
	{title} {\bibinfo {title} {{Phases, transitions, and patterns in the
				one-dimensional Extended Bose-Hubbard model}},\ }\href
	{http://arxiv.org/abs/1403.2315} {\  (\bibinfo {year} {2014})},\ \Eprint
	{https://arxiv.org/abs/1403.2315} {arXiv:1403.2315} \BibitemShut {NoStop}%
	\bibitem [{\citenamefont {Yang}(1962)}]{Yang1962}%
	\BibitemOpen
	\bibfield  {author} {\bibinfo {author} {\bibfnamefont {C.~N.}\ \bibnamefont
			{Yang}},\ }\bibfield  {title} {\bibinfo {title} {{Concept of Off-Diagonal
				Long-Range Order and the Quantum Phases of Liquid He and of
				Superconductors}},\ }\href {https://doi.org/10.1103/RevModPhys.34.694}
	{\bibfield  {journal} {\bibinfo  {journal} {Rev. Mod. Phys.}\ }\textbf
		{\bibinfo {volume} {34}},\ \bibinfo {pages} {694} (\bibinfo {year}
		{1962})}\BibitemShut {NoStop}%
	\bibitem [{\citenamefont {Lewenstein}\ \emph {et~al.}(2012)\citenamefont
		{Lewenstein}, \citenamefont {Sanpera},\ and\ \citenamefont
		{Ahufinger}}]{Lewenstein}%
	\BibitemOpen
	\bibfield  {author} {\bibinfo {author} {\bibfnamefont {M.}~\bibnamefont
			{Lewenstein}}, \bibinfo {author} {\bibfnamefont {A.}~\bibnamefont
			{Sanpera}},\ and\ \bibinfo {author} {\bibfnamefont {V.}~\bibnamefont
			{Ahufinger}},\ }\href
	{https://doi.org/10.1093/acprof:oso/9780199573127.001.0001} {\emph {\bibinfo
			{title} {{Ultracold Atoms in Optical Lattices}}}}\ (\bibinfo  {publisher}
	{Oxford University Press},\ \bibinfo {address} {Oxford},\ \bibinfo {year}
	{2012})\BibitemShut {NoStop}%
	\bibitem [{\citenamefont {Xiao-Gang}(2007)}]{XGWen}%
	\BibitemOpen
	\bibfield  {author} {\bibinfo {author} {\bibfnamefont {W.}~\bibnamefont
			{Xiao-Gang}},\ }\href
	{https://doi.org/10.1093/acprof:oso/9780199227259.001.0001} {\emph {\bibinfo
			{title} {{Quantum field theory of many-body systems: from the origin of sound
					to an origin of light and electrons}}}}\ (\bibinfo  {publisher} {Oxford
		University Press},\ \bibinfo {address} {Oxford},\ \bibinfo {year}
	{2007})\BibitemShut {NoStop}%
	\bibitem [{\citenamefont {Calabrese}\ and\ \citenamefont
		{Cardy}(2004)}]{Calabrese2004}%
	\BibitemOpen
	\bibfield  {author} {\bibinfo {author} {\bibfnamefont {P.}~\bibnamefont
			{Calabrese}}\ and\ \bibinfo {author} {\bibfnamefont {J.}~\bibnamefont
			{Cardy}},\ }\bibfield  {title} {\bibinfo {title} {{Entanglement entropy and
				quantum field theory}},\ }\href
	{https://doi.org/10.1088/1742-5468/2004/06/P06002} {\bibinfo  {journal} {J.
			Stat. Mech.: Theory Exp.}\ ,\ \bibinfo {pages} {P06002}}\BibitemShut
	{NoStop}%
	\bibitem [{\citenamefont {Mikeska}\ and\ \citenamefont
		{Kolezhuk}(2004)}]{Mikeska2005}%
	\BibitemOpen
	\bibfield  {journal} {  }\bibfield  {author} {\bibinfo {author} {\bibfnamefont
			{H.~J.}\ \bibnamefont {Mikeska}}\ and\ \bibinfo {author} {\bibfnamefont
			{A.~K.}\ \bibnamefont {Kolezhuk}},\ }\bibfield  {title} {\bibinfo {title}
		{{One-Dimensional Magnetism}},\ }\href {https://doi.org/10.1007/bfb0119591}
	{\bibfield  {journal} {\bibinfo  {journal} {Lect. Notes Phys.}\ }\textbf
		{\bibinfo {volume} {645}},\ \bibinfo {pages} {1} (\bibinfo {year}
		{2004})}\BibitemShut {NoStop}%
	\bibitem [{\citenamefont {Kennedy}\ and\ \citenamefont
		{Tasaki}(1992)}]{KennedyTasaki}%
	\BibitemOpen
	\bibfield  {author} {\bibinfo {author} {\bibfnamefont {T.}~\bibnamefont
			{Kennedy}}\ and\ \bibinfo {author} {\bibfnamefont {H.}~\bibnamefont
			{Tasaki}},\ }\bibfield  {title} {\bibinfo {title} {{Hidden symmetry breaking
				and the Haldane phase inS=1 quantum spin chains}},\ }\href
	{https://doi.org/10.1007/BF02097239} {\bibfield  {journal} {\bibinfo
			{journal} {Commun. Math. Phys.}\ }\textbf {\bibinfo {volume} {147}},\
		\bibinfo {pages} {431} (\bibinfo {year} {1992})}\BibitemShut {NoStop}%
	\bibitem [{\citenamefont {Affleck}\ \emph {et~al.}(1988)\citenamefont
		{Affleck}, \citenamefont {Kennedy}, \citenamefont {Lieb},\ and\ \citenamefont
		{Tasaki}}]{AKLT}%
	\BibitemOpen
	\bibfield  {author} {\bibinfo {author} {\bibfnamefont {I.}~\bibnamefont
			{Affleck}}, \bibinfo {author} {\bibfnamefont {T.}~\bibnamefont {Kennedy}},
		\bibinfo {author} {\bibfnamefont {E.~H.}\ \bibnamefont {Lieb}},\ and\
		\bibinfo {author} {\bibfnamefont {H.}~\bibnamefont {Tasaki}},\ }\bibfield
	{title} {\bibinfo {title} {{Valence bond ground states in isotropic quantum
				antiferromagnets}},\ }\href {https://doi.org/10.1007/BF01218021} {\bibfield
		{journal} {\bibinfo  {journal} {Commun. Math. Phys.}\ }\textbf {\bibinfo
			{volume} {115}},\ \bibinfo {pages} {477} (\bibinfo {year}
		{1988})}\BibitemShut {NoStop}%
	\bibitem [{\citenamefont {Soriano}\ and\ \citenamefont
		{Palacios}(2014)}]{Soriano2014}%
	\BibitemOpen
	\bibfield  {author} {\bibinfo {author} {\bibfnamefont {M.}~\bibnamefont
			{Soriano}}\ and\ \bibinfo {author} {\bibfnamefont {J.~J.}\ \bibnamefont
			{Palacios}},\ }\bibfield  {title} {\bibinfo {title} {{Theory of projections
				with nonorthogonal basis sets: Partitioning techniques and effective
				Hamiltonians}},\ }\href {https://doi.org/10.1103/PhysRevB.90.075128}
	{\bibfield  {journal} {\bibinfo  {journal} {Phys. Rev. B}\ }\textbf {\bibinfo
			{volume} {90}},\ \bibinfo {pages} {075128} (\bibinfo {year}
		{2014})}\BibitemShut {NoStop}%
	\bibitem [{\citenamefont {Toeplitz}(1911)}]{Toeplitz1911}%
	\BibitemOpen
	\bibfield  {author} {\bibinfo {author} {\bibfnamefont {O.}~\bibnamefont
			{Toeplitz}},\ }\bibfield  {title} {\bibinfo {title} {{Zur Theorie der
				quadratischen und bilinearen Formen von unendlichvielen
				Ver{\"{a}}nderlichen}},\ }\href@noop {} {\bibfield  {journal} {\bibinfo
			{journal} {Math. Ann.}\ }\textbf {\bibinfo {volume} {70}},\ \bibinfo {pages}
		{351} (\bibinfo {year} {1911})}\BibitemShut {NoStop}%
	\bibitem [{\citenamefont {Auerbach}(1994)}]{assa}%
	\BibitemOpen
	\bibfield  {author} {\bibinfo {author} {\bibfnamefont {A.}~\bibnamefont
			{Auerbach}},\ }\href@noop {} {\emph {\bibinfo {title} {{Interacting Electrons
					and Quantum Magnetism}}}}\ (\bibinfo  {publisher} {Springer-Verlag},\
	\bibinfo {address} {Berlin},\ \bibinfo {year} {1994})\BibitemShut {NoStop}%
\end{thebibliography}
%

\end{document}